\documentclass[12pt]{iopart}

\usepackage{cite}
\usepackage{enumerate}
\usepackage{amssymb}
\usepackage{amsbsy}
\usepackage[dvips]{graphicx}
\usepackage{color}
\usepackage{graphics}

\newcommand{\be}{\begin{equation}} \newcommand{\ee}{\end{equation}}
\newcommand{\bea}{\begin{eqnarray}} \newcommand{\eea}{\end{eqnarray}}
\newcommand{\el}{\nonumber \\}
\newcommand{\re}[1]{(\ref{#1})}
\newcommand{\bfm}{\mathbf}
\newcommand{\pat}{\partial}
\newcommand{\abs}[1]{|#1|}
\renewcommand{\sec}[1]{section \ref{#1}}
\newcommand{\fig}[1]{figure \ref{#1}}
\newcommand{\brt}[1]{[#1]}

\newcommand{\mpc}{\mbox{$h^{-1}$Mpc}}
\newcommand{\LCDM}{$\Lambda$CDM\ }

\newcommand{\adot}{\dot{a}}
\newcommand{\addot}{\ddot{a}}
\newcommand{\rhodot}{\dot{\rho}}

\newcommand{\deltadot}{\dot{\delta}}
\newcommand{\deltaddot}{\ddot{\delta}}
\newcommand{\bx}{\bi{x}}

\renewcommand{\H}{\frac{\adot}{a}}
\newcommand{\HH}{\frac{\adot^2}{a^2}}
\newcommand{\av}[1]{\langle{#1}\rangle}
\newcommand{\sQ}{\mathcal{Q}}
\newcommand{\sR}{{^{(3)}R}}
\newcommand{\Om}{\Omega_m}
\newcommand{\Omn}{\Omega_{m0}}
\newcommand{\OQ}{\Omega_{\sQ}}
\newcommand{\OQn}{\Omega_{\sQ0}}
\newcommand{\OR}{\Omega_{R}}
\newcommand{\ORn}{\Omega_{R0}}

\newcommand{\PRD}[1]{{\it Phys. Rev.} {\bf D#1}}
\renewcommand{\PRL}[1]{{\it Phys. Rev. Lett.} {\bf #1}}

\newcommand{\NPB}[1]{{\it Nucl. Phys.} {\bf B#1}}
\newcommand{\PLA}[1]{{\it Phys. Lett.} {\bf A#1}}
\newcommand{\PLB}[1]{{\it Phys. Lett.} {\bf B#1}}
\newcommand{\MNRAS}[1]{{\it Mon. Not. Roy. Astron. Soc.} {\bf #1}}
\newcommand{\APJ}[1]{{\it Astrophys. J.} {\bf #1}}
\newcommand{\APJS}[1]{{\it Astrophys. J. Suppl.} {\bf #1}}

\renewcommand{\CQG}[1]{{\it Class. Quant. Grav.} {\bf #1}}
\newcommand{\GRG}[1]{{\it Gen. Rel. Grav.} {\bf #1}}
\renewcommand{\AA}[1]{{\it Astron. \& Astrophys.} {\bf #1}}
\newcommand{\PROG}[1]{{\it Prog. Theor. Phys.} {\bf #1}}
\newcommand{\AJ}[1]{{\it Astron. J.} {\bf #1}}

\begin{document}

\begin{titlepage}

\begin{flushleft}
  \hfill            CERN-PH-TH/2006-148 \\ \hfill \\
\end{flushleft}

\title{Accelerated expansion from structure formation}

\author{Syksy R\"{a}s\"{a}nen}

\address{CERN, Physics Department Theory Unit, CH-1211 Geneva 23, Switzerland}

\ead{syksy {\it dot} rasanen {\it at} iki {\it dot} fi}

\begin{abstract}

\noindent
We discuss the physics of backreaction-driven accelerated expansion.
Using the exact equations for the behaviour of averages in dust universes,
we explain how large-scale smoothness does not imply that the effect of
inhomogeneity and anisotropy on the expansion rate is small.
We demonstrate with an analytical toy model how gravitational collapse
can lead to acceleration. We find that the conjecture of the accelerated
expansion being due to structure formation is in agreement with the
general observational picture of structures in the universe, and
more quantitative work is needed to make a detailed comparison.

\end{abstract}

\pacs{04.40.Nr, 95.36.+x, 98.80.-k, 98.80.Jk}

\end{titlepage}

\tableofcontents

\setcounter{secnumdepth}{3}

\section{Introduction} \label{sec:intro}

\paragraph{Evidence for acceleration.}

There is a large body of observational evidence supporting
the claim that the expansion of the universe has accelerated
in the recent past, and may be accelerating today.
This conclusion has been bolstered by the verification of the
prediction of the location of the baryon acoustic peak in the matter
power spectrum \cite{BAO}, in a convincing demonstration of concordance.
In addition, the worrisome feature
that nearby and distant populations of type Ia supernovae
used to have different absolute magnitudes and
were both individually consistent with deceleration
has disappeared with new and better data \cite{Padmanabhan}
(though see \cite{Shapiro:2005, Elgaroy:2006}).

The \LCDM model, where the acceleration is driven
by vacuum energy (or the cosmological constant, which is
the equivalent modification of gravity) agrees well with most
observations, with the notable exception of the low CMB multipoles
\cite{CMBmaps, Hansen:2004, Bielewicz:2005, dipolesyst, Copi:2006}
(it has also been argued that cluster observations support a
non-accelerating universe \cite{Blanchard}).
However, given the lack of theoretical understanding about
the parameters of the \LCDM model (notably the vacuum energy density),
it is a phenomenological fit rather than a well-founded theory,
and its success does not rule out the possibility that quite
a different model can also be a good fit to the data.
(The values obtained for the parameters of a cosmological model
by fitting to observations should not be mistaken for measurements,
as model selection studies show; see e.g.
\cite{Mukherjee:2005, Shapiro:2005, Elgaroy:2006}.)
In particular, while the observation that there is accelerating
expansion seems
robust, the nature of the acceleration is not well constrained.
In the \LCDM model, the transition to acceleration is gradual,
but a rapid transition is not ruled out \cite{trans, Ichikawa:2006}.
In fact, from the SNIa data it is difficult to say anything beyond
that the universe has accelerated in the recent past, even
whether the expansion is still accelerating
\cite{Shapiro:2005, Elgaroy:2006, Gong:2006}.

Keeping to the assumption that the universe is completely
homogeneous and isotropic, any explanation of the acceleration
has to involve either a medium with negative pressure or
modified gravity. Such models in general, and the \LCDM
model in particular, suffer from {\it the coincidence problem}:
why does the acceleration happen around a redshift of
unity, at around 10 billion years? In other words, why are we
seeing a very particular phase in the evolution of the universe,
when the inferred energy density of the source driving the acceleration
has recently become equal to the energy density of matter?
The clearest qualitative change in the late-time universe
is the formation of non-linear structures. It therefore seems a natural
possibility that the observed deviation from the prediction of
homogeneous and isotropic cosmological models with normal matter
and gravity could be related to the known breakdown of the assumption
that the universe is homogeneous and isotropic (rather than to a
speculated failure in the description of the matter content or
the theory of gravity).

\paragraph{The inhomogeneous universe.}

One possible avenue is trying to explain the observations
without having any accelerated expansion.
Cosmological information is borne to us by light along null
geodesics (apart from information carried by neutrinos and cosmic rays).
The standard analysis of light propagation assumes that 
the universe is perturbatively near a homogeneous and
isotropic Friedmann--Robertson--Walker (FRW) model, which
is manifestly not true on scales smaller than 70-100 \mpc
\cite{Hogg:2004, Pietronero, morphology}.

It is therefore possible that the propagation of light would be
affected by the inhomogeneities and/or anisotropies in a way
that looks like acceleration when interpreted in the context of
an FRW model. Studies of the Lema\^{\i}tre--Tolman--Bondi (LTB)
model \cite{LTB} (see \cite{Krasinski:1997} for a review),
the spherically symmetric dust solution of the Einstein
equation, have demonstrated that the effect of inhomogeneity
on the luminosity distance can mimic acceleration
\cite{LTBgeo, Biswas:2006} (see \cite{Biswas:2006} for more references).
Even though spherical symmetry is a questionable assumption
for the entire universe, it could be a good first
approximation for the local region. In any case, one would expect
a qualitatively similar effect to be present also in more
realistic and less symmetric spacetimes \cite{Bolejko}
-- arguably, the effect of clumpiness could even be stronger
when there is less symmetry.
The effect of inhomogeneity and anisotropy on the luminosity
distance has also been studied in perturbed FRW models
\cite{Barausse:2005, Bonvin}.

An explanation of the apparent acceleration in terms of
inhomogeneity and/or anisotropy could solve the coincidence problem,
since inhomogeneity and anisotropy become important only in the
late-time universe.
However, inhomogeneity and anisotropy affect different observations in
different ways, and it would require an odd coincidence for all
the various indicators of expansion rate (SNIa luminosity distances,
the cosmological microwave background (CMB) anisotropies, large
scale structure (LSS), and so on) to be affected in a way that
would be consistently interpreted as acceleration when fitting to a FRW model.

One proposed possibility is that we live in an
underdense region, a 'Hubble bubble' \cite{hubbub}
(for more references, see \cite{Biswas:2006}).
In this proposal, the local matter density today is
$\Omn\approx$ 0.15-0.35, as indicated by local observations
\cite{Peebles:2004}, while the global value is $\Om=1$.
The SNIa luminosity distances as well as the difference
between the local and global values of the expansion rate could
be explained in terms of inhomogeneity, while a global model with no
acceleration can fit most other observations, including the CMB and LSS
\cite{Sarkar} (though it is not supported by studies of the
local and global expansion rate \cite{Sandage:2006}).
However, it would be difficult to explain the baryon
acoustic peak \cite{Blanchard:2005}: one would have to appeal
to inhomogeneities (or features in the primordial power spectrum)
to supply a pattern that by coincidence happens to fit the expectations
of an accelerating FRW model.

One way to phrase the issue is that cosmological observations
involve a larger number of a priori independent parameters than the
\LCDM model. Therefore the \LCDM model implies relations between
observationally independent parameters.
(For an early discussion of cosmological observations in an
inhomogeneous and anisotropic spacetime which makes the issue
transparent, see \cite{Kristian:1966}.)
It is not surprising that models with more degrees of freedom,
such as the LTB model which involves two arbitrary functions,
could fit the data as well as FRW models. However, it would
be an unlikely coincidence for them to also produce the
same relations between observables as FRW models.
This is generally true for any  models where the explanation
of the luminosity distances is decoupled from the explanation
of the low matter density, baryon acoustic peak and so on
(such as mixing of photons with axions \cite{Csaki:2001} or with the
gauge bosons of a new $U(1)$ gauge group \cite{Evslin:2005}).

\paragraph{The fitting problem.}

While the FRW scale factor has been very
successful in fitting observations, it is
difficult to understand the matter content implied by
the FRW equations which relate the scale factor to
the energy--momentum tensor.
Note that the fact that the mean properties of the universe
are well described by an overall scale factor does not imply the
stronger statement that the scale factor evolves according to
the FRW equations, since the universe is not completely homogeneous
and isotropic.

The idea that the average behaviour of inhomogeneous and/or
anisotropic spacetimes is in general different from the
behaviour of homogeneous and isotropic spacetimes
goes back to at least 1963 \cite{Shirokov:1963}. The
first comprehensive discussion was given in 1983 by George Ellis
\cite{fitting}, who called the task of finding the smooth metric
which best fits the real clumpy universe {\it the fitting problem}.
The influence of inhomogeneity and/or anisotropy on the average behaviour
is also known as backreaction
\cite{Kasai, Buchert:1995, Woodard, Ehlers:1996, Unruh:1998, Takada:1999, Buchert:1999, Sicka, Taruya:1999, Buchert:2000, Kerscher:2000, Carfora, Buchert:2001, Tatekawa:2001, Geshnizjani:2002, Schwarz, Buchert:2002, Brandenberger:2002, Geshnizjani:2003, Rasanen:2003, Buchert:2003, Hosoya:2004, Rasanen:2004, Kolb:2004a, Kolb:2004b, Barausse:2005, Kolb:2005a, Flanagan:2005, Geshnizjani:2005, Hirata:2005, Notari:2005, Rasanen:2005, Siegel:2005, Martineau:2005a, Tsamis:2005, Kolb:2005b, global, Ishibashi:2005, Losic, Martineau:2005b, Kolb:2005c, Kolb:2005d, Chuang:2005, Kasai:2006, Parry:2006, Kai:2006, Paranjape:2006, Rasanen:2006, Buchert:2006};
see \cite{Krasinski:1997, Rasanen:2003} for further references, in
particular early ones, and \cite{Ellis:2005} for an overview.

The idea that perturbations with wavelengths
smaller than the Hubble radius could lead to acceleration for the
scale factor (as opposed to merely mimicking the appearance of acceleration
via changing null geodesics) was studied in the context of linear
perturbation theory in \cite{Rasanen:2003} (the possibility had been
earlier touched upon in \cite{Buchert:2000, Tatekawa:2001};
see also \cite{Schwarz}).
The calculation was then extended to second order
\cite{Barausse:2005, Kolb:2004a}, and it was suggested that
linear perturbations with wavelengths much larger than the Hubble
radius could lead to acceleration
\cite{Barausse:2005, Kolb:2004a, Kolb:2004b, Kolb:2005a}.
It is now agreed that this is not possible
\cite{Flanagan:2005, Geshnizjani:2005, Hirata:2005, Rasanen:2005, Kolb:2005b}\footnote{Even though super-Hubble perturbations do not contribute
to acceleration in a dust universe, they could lead to deceleration
during inflation driven by a scalar field or a cosmological constant,
and super-Hubble scalar field perturbations left over from inflation
could still be important today
\cite{Woodard, Unruh:1998, Geshnizjani:2002, Brandenberger:2002, Geshnizjani:2003, Martineau:2005a, Tsamis:2005, Losic, Martineau:2005b, Kolb:2005d}.}.
As for perturbations smaller than the Hubble radius,
there is no acceleration to at least second order in perturbation theory
\cite{Rasanen:2003, Kolb:2004a, Kolb:2005b, Kasai:2006, Parry:2006}.
If inhomogeneity and anisotropy are to explain the observed acceleration,
the only possibility is via non-linear sub-Hubble perturbations,
that is, the process of structure formation, as proposed in
\cite{Schwarz, Rasanen:2003}.

It has been analytically shown in the LTB toy model how
backreaction of non-linear perturbations
can modify the Hubble law \cite{Rasanen:2004}, and
acceleration has also been numerically demonstrated in
the LTB model \cite{Chuang:2005, Kai:2006, Paranjape:2006},
but the physical meaning of inhomogeneity- and anisotropy-driven
acceleration and the connection to structure formation has been unclear.
We will discuss the relation between homogeneity and isotropy,
the overall scale factor, the FRW metric and the equations
which describe the average expansion of the universe.
We will then look at an exact toy model of structure formation,
explain the physics of acceleration driven by inhomogeneity
and anisotropy, and note that structure formation involves a preferred
time near the observed acceleration era. In particular, we will
clarify two apparent paradoxes of backreaction-driven acceleration:
how the average expansion of a manifold can accelerate even though
the local expansion rate decelerates everywhere, and how collapse
implies acceleration. These issues were earlier discussed in the
brief essay \cite{Rasanen:2006}.

In \sec{sec:smoo} we discuss the assumptions underlying FRW
models, and go through the derivation of the Buchert equations which
describe the average behaviour of an inhomogeneous
and/or anisotropic dust spacetime.
In \sec{sec:acc} we consider an exact toy model where
gravitational collapse produces acceleration, discuss how
this mechanism may operate in the real universe, and compare
this picture with some observations and simulations of
structures in the universe.
In \sec{sec:con} we summarise the situation with regard to
the conjecture that the observed acceleration is due to backreaction.

\section{Smoothness and variance} \label{sec:smoo}

\subsection{The Friedmann--Robertson--Walker assumptions} \label{sec:FRW}

\paragraph{The three assumptions.}

The assumption of homogeneity and isotropy that underlies the
Friedmann--Robertson--Walker models of cosmology can be broken down
into three distinct parts.

\begin{enumerate}[{Assumption} 1:]

\item {\it FRW scale factor.}
The observables characterising the mean properties of the universe
can be computed from an overall scale factor.

\item {\it FRW dynamics.}
The overall scale factor evolves according to the FRW equations.

\item  {\it FRW + perturbations.}
Deviations from homogeneity and isotropy evolve according
to linear perturbation theory around the average.

\end{enumerate}

\noindent Since the universe is not exactly homogeneous and
isotropic, it is clear that the above assumptions do not hold
exactly. However, there is usually no attempt to quantify the
deviation, and the first two assumptions are often conflated.
A notable exception is the program of observational cosmology
(and related work) by George Ellis and collaborators,
which aimed at formulating cosmological theory in a manner that
does not involve a priori assumptions about homogeneity
and isotropy and that is as close to the observations as possible
\cite{Ellis:1975, obscos, hom, Matravers:1995};
see \cite{Ellis:1999, Ellis:2000} for an overview.

We will look at the above assumptions from the slightly different
point of view of backreaction studies, where the emphasis is not so
much on establishing the level of inhomogeneity and isotropy of the
universe as on quantifying their effect on the expansion rate and
other observables. Let us discuss the three assumptions in turn.

\paragraph{Assumption 1: FRW scale factor.}

Observations indicate that the universe is statistically
homogeneous and isotropic on large scales.
It has recently become possible to directly establish
from the Sloan Digital Sky Survey, for the first time,
the average homogeneity of the universe by looking at the
fractal dimension of the point set of luminous red galaxies
\cite{Hogg:2004, Pietronero}. The related homogeneity scale has
been quantified as 70-100 \mpc, though analysis based on the
morphology of structures indicates a homogeneity scale that is
larger than 100-200 \mpc \cite{morphology}.
Dividing the observational volume into 10 regions with
individual volume 2$\times10^7$ (\mpc)$^3$ (corresponding to a
ball with radius $\approx$ 170 \mpc), the density variance is 7\%
in the redshift range $0.2<z<0.35$, quantifying the
degree of statistical homogeneity in present observations.
The largest structure known is the Sloan Great Wall at 420 \mpc,
which has superseded the old 240 \mpc\ Great Wall.
These sizes are 14\% and 8\%, respectively, of the Hubble
radius; in the Einstein-de Sitter universe (the spatially flat
matter-dominated FRW model) this would be 7\% and 4\%,
respectively, of the visual horizon.
Structures this large are rare, and the typical size
of observed collapsing structures and voids is
$\approx$ 20-40 \mpc, of the order $10^{-3}$ to $10^{-2}$ of the visual
horizon \cite{voidobs, Hoyle:2001, Hoyle:2003} (though some simulations
suggest that a significant fraction would be larger \cite{Shandarin}
or smaller \cite{Colberg:2004}).

Statistical isotropy is in turn supported by the 
high degree of isotropy in the CMB, along with the
'almost Ehlers--Geren--Sachs theorem', which states that a
universe where the CMB looks almost isotropic everywhere is
almost FRW on large scales \cite{Stoeger:1995, EGS} (see
\cite{Wainwright, Clarkson} for discussion and caveats).
The applicability of the 'almost EGS theorem' to the real
universe is, in fact, somewhat unclear.
The proof of the theorem requires the assumption that the expansion
rate is positive everywhere, which is not valid in the real
universe unless scales where structure formation by gravitational
collapse is relevant are smoothed over.
Even then, it is unclear how the strict limits
(essentially given by the CMB anisotropy of $10^{-5}$)
on spatial variation of the exansion rate and other
observables can be reconciled with the observed (and theoretically
expected) differences of order one in the expansion rate and energy
density in non-linear regions (we will discuss the observations in
\sec{sec:variance}). Nevertheless, it seems intuitively clear
that the isotropy of the CMB (coupled with the assumption that
we do not occupy a preferred position in the universe) indicates
a high degree of average isotropy in the geometry on large scales.

Statistical homogeneity and isotropy show that a description in
terms of an overall scale factor could be consistent.
In physical terms, because the size of typical non-linear
regions is small, light rays coming to us cosmological distances
pass through several such regions on the way to us, and the
differences could be expected to average out.
However, the small size of non-linear regions does not prove that
a description in terms of an overall scale factor is necessarily correct.
We receive almost all cosmological information along null geodesics,
and from the fact that inhomogeneities and anisotropies are small
when averaged over large scales it does not follow that a description
in terms of an overall scale factor will correctly capture the
physics of light propagation, as discussed in \sec{sec:intro}.

The FRW metric is conformally flat, so the Weyl tensor
which embodies the non-local effects of gravity vanishes, while
the Ricci tensor, determined by the local matter distribution, is non-zero.
However, photons in the real universe only occasionally encounter
matter and mostly travel in vacuum, where the Weyl tensor is non-zero
but the Ricci tensor vanishes (neglecting the small contribution of the
microwave and neutrino backgrounds).
The FRW description is exactly the opposite of the real situation,
and it is not obvious why it would correctly capture the physics
of the passage of light in the real inhomogeneous and anisotropic
universe. From a theoretical point of view, this issue has not been
satisfactorily settled \cite{Ellis:1998, Lieu:2004a, Kibble:2004, Lieu:2004b};
for an overview, see \cite{Ellis:2005} and for further references,
see \cite{Ellis:2005, Biswas:2006}.

A related issue is that cosmological models are usually
discussed in terms of hypersurfaces of proper time, whereas
observations are mostly made along the past lightcone.
This issue was considered in detail in the program of
observational cosmology \cite{obscos}.
In FRW models this distinction is not important, and they enjoy
(for monotonous expansion) a simple one-to-one correspondence
between time and redshift. In an inhomogeneous and/or anisotropic
universe, the issue is more complicated, and is related to the
problem of choosing the hypersurface of proper time
\cite{Geshnizjani:2002, Geshnizjani:2003, Rasanen:2003, Rasanen:2004}.
Looking only at an average scale factor has been criticised in
\cite{Ishibashi:2005} (see also \cite{Kolb:2005c}) on the grounds
that one obtains different behaviour for different choices of time
slicing.

In FRW models with only a single fluid, there is a preferred
time coordinate given by the proper time measured by observers
comoving with the fluid. This notion can be straightforwardly
extended to inhomogeneous and/or anisotropic dust
spacetimes which do not have any symmetries. If the matter consists
of irrotational dust, then such hypersurfaces of constant proper time
are everywhere orthogonal to the fluid flow lines, and the situation
is analogous to the FRW case.
If vorticity is present, the hypersurfaces of proper time will not
mesh together to fill the spacetime \cite{Ehlers:1961, Ellis:1971},
and the situation is more complicated
\cite{Hirata:2005, Rasanen:2005, Stoeger:1995}.
(During inflation driven by a cosmological constant or a scalar field,
the issue of time slicing and observables is more involved
\cite{Unruh:1998, Geshnizjani:2002, Geshnizjani:2003, Tsamis:2005}.)
In any case, averaging does involve a loss of covariance; see
\cite{Ellis:1995} for discussion.

The approximation that the matter is an irrotational,
pressureless ideal fluid will necessarily break down at small
scales once structure formation has started
\cite{DelPopolo:2001, Buchert:2005a}, as otherwise collapsing
structures could not stabilise, given that the local expansion
rate of irrotational dust can never increase.
In other words, structure formation and the associated vorticity need
to be considered and may be important for backreaction.
As we will see, zero vorticity is an important mathematical assumption
for the definition of averages and an overall scale factor.
However, from a physical point view one would not expect the
vorticity associated with structure formation to make a quantitative
difference for the expansion of the universe.
The issue should be carefully considered, but we shall simply
assume that the small-scale breakdown of the picture of matter
as an irrotational ideal fluid with zero pressure is not important.

Note that the problem of time and the averaging hypersurface
is not an artifact of inhomogeneous and/or anisotropic models,
but a feature of the general relatistic description of the real universe.
It is a virtue of the observational cosmology and backreaction
approaches that they make these issues explicit and
provide tools for studying and quantifying them, unlike FRW models.

A seemingly worrisome aspect of the hypersurfaces of constant proper
time is that observables averaged over them depend on regions outside
the visual horizon, since the hypersurfaces extend beyond the past
lightcone. It may seem unphysical that an observational
quantity would change if we changed the definition of the hypersurface
in regions beyond our past lightcone. However, if the universe is
statistically homogeneous and isotropic on large scales, we do
not have the freedom to independently adjust the hypersurface
of proper time in the regions inside and outside our past lightcone,
since the state of matter and geometry inside and outside the lightcone is
required to be statistically identical for the same proper time.
(See \cite{Ellis:1975, hom} for discussion of homogeneity in cosmology.)
This puzzle is therefore just the homogeneity and isotropy problem,
which is solved (or at least alleviated \cite{Trodden}) by inflation.
The statistical equality of widely separated regions was set up
in the early universe when they were in causal contact, so
the particle horizon is much larger than the visual horizon.
(For clarification of the different horizons, see
\cite{Davis:2003, Ellis:2006}.)
In spacetimes which are not statistically homogeneous and
isotropic, the averaging procedure would not be
expected to be useful, and a description in terms of
an overall scale factor would probably not make sense.
For example, this would be the case if the size of
typical non-linear structures were a sizeable fraction of
the visual horizon.

We will simply extend the FRW notions of proper time and
scale factor in the most straightforward manner (as we will discuss
in section \ref{sec:Buchert}) and assume that redshift is related
to the scale factor in the same way as in FRW models.
Dependence on proper time could then be expressed in terms
of redshift, as in the FRW case, so that the problem
discussed above is not apparent. Of course, this is simply a matter
of rewriting the equations in a manner that makes the assumption of
statistical homogeneity and isotropy less transparent.

Properly, one should derive the quantitive conditions under which
the scale factor approximation is valid and see to which extent they are
realised in the universe. This would involve defining redshifts,
luminosity distances and other observables using null geodesics
in the inhomogeneous and anisotropic universe, and determining
under which conditions
these reduce to the quantities defined with the overall scale factor.
As discussed above, the issue of average light propagation
in inhomogeneous and/or anisotropic spacetimes has not been
satisfactorily settled from the theoretical point of view.

Nevertheless, from an observational point of view, the scale factor
ansatz has been very successful in fitting a range of observations.
This success is all the more remarkable as different observations
depend on null geodesics in a different manner.
For example, there are several different definitions of
the expansion rate (such as the volume expansion rate, the rate of
deviation between neighbouring geodesics, the expansion rate inferred
from the luminosity distance and so on), which agree for a homogeneous
and isotropic space but differ when inhomogeneities
and/or anisotropies are present.
Yet, expansion rates inferred from different observations agree
to within $\approx20\%$ \cite{Sandage:2006}.
The approximation of considering only the scale factor has worked
very well in practice, whatever its theoretical status, and the
observed statistical homogeneity and isotropy on large scales at least
guarantees that this approximation is consistent.

Note that even small corrections to the approximation of looking
only at the scale factor can be interesting for the light they
shed on the average behaviour. For example, the
dipole of the angular power spectrum of the inhomogeneous luminosity
distance in a linearly perturbed FRW model yields a direct measure of
the Hubble parameter as a function of redshift \cite{Bonvin}.
Also, even if the passage of light through cosmological distances,
and thus many non-linear regions, is well described in terms
of a scale factor, this does not rule out the possibility that
local structures could affect null geodesics in a way that is
not captured by that approximation.
In particular, the anomalies of the low CMB multipoles
\cite{CMBmaps, Hansen:2004, Bielewicz:2005, dipolesyst, Copi:2006}
could be related to the local breakdown of the description
in terms of a linearly perturbed FRW metric \cite{CMBmodels1, CMBmodels2}.

(Another distinct issue, which we will not discuss, is the ``dressing''
of cosmological parameters, i.e. the feature that the usual interpretation
of observations does not account for the geometrical inhomogeneities
and/or anisotropies in the averaging domain when considering observables
\cite{Carfora, Buchert:2002, Buchert:2003}.)

Even if the effects of inhomogeneity and anisotropy cancel
on large scales so that the FRW {\it scale factor} is a good
average description, this does not mean that the FRW {\it metric}
would be a good approximation of the average geometry.
The reason is that the FRW metric also contains the
spatial curvature, which is assumed to be homogeneous and isotropic.
In a general spacetime, the spatial curvature is inhomogeneous
and anisotropic, and does not evolve like the FRW
spatial curvature, even on average.
The reason is that the evolution of non-linear regions with
positive spatial curvature and those with negative spatial curvature
is different, and they are not correlated so as to produce the
FRW behaviour.
In particular, even if a universe is perturbatively close to
spatial flatness early on, this condition is not necessarily
preserved once density perturbations become non-linear.

In mathematical terms, this is related to the fact that the FRW
evolution $\propto a^{-2}$ of the spatial curvature in terms of
the scale factor $a$ arises from the integrability condition between
the Raychaudhuri equation and the Hamiltonian constraint. The two receive
different corrections from the inhomogeneities and anisotropies,
so the general integrability condition does not agree with the FRW case and
the average spatial curvature does not in general evolve like $a^{-2}$.
From a physical point of view, regions with
negative spatial curvature expand faster than regions with positive
spatial curvature, so one would expect that they will come to dominate
the volume and the average spatial curvature will become negative.
The non-FRW evolution of the spatial curvature and the competition
between overdense and underdense regions is central to the proposed
backreaction explanation for accelerated expansion, and we will
discuss these issues in detail in sections \ref{sec:Buchert} and \ref{sec:toy}.
The behaviour of spatial curvature is at the heart of the distinction
between the scale factor being a good description and the scale
factor following the FRW equations, an issue to which we now turn.

\paragraph{Assumption 2: FRW dynamics.}

As discussed above, we will simply assume that the average
properties of the universe can be described in terms of an
overall scale factor, without deriving the conditions under
which this assumption is valid. In contrast, when considering
the dynamics, we will write down the exact equations governing
the evolution of the scale factor, quantify the domain of validity
of the FRW equations and discuss the impact of the corrections.
Indeed, while there is strong observational support for the
FRW scale factor, the time evolution
given by the FRW equations is quite different from what is observed,
unless the equations are amended by adding a source term with
negative pressure or by modifying gravity.

It has been suggested that in order to emphasise the difference
between the geometry and the dynamics, the names Robertson--Walker
would be associated with the assumption that the geometry is
approximately homogeneous and isotropic, while the stronger
assumption that the dynamics of the scale factor is given by
the Einstein equation applied to a completely homogeneous and
isotropic metric would bear the names
Friedmann--Lema\^{\i}tre \cite{Ellis:1999, Ellis:2000}.
(As discussed above, the present situation is slightly different
in that we are not even assuming that the geometry is described
by a homogeneous and isotropic metric, simply that we can use an
overall scale factor.)

Simply inserting the overall scale factor into the Einstein
equation is not the correct way to find the dynamical equations
which it satisfies.
Instead, one should insert the full inhomogeneous and/or anisotropic
metric into the Einstein equation and only then take an average, since
the average behaviour of an inhomogeneous and/or anisotropic spacetime
is not the same as the behaviour of the corresponding smooth
spacetime (where ``corresponding'' means that the smooth
and average quantities have the same initial conditions).
In other words, the average properties of an inhomogeneous
and/or anisotropic spacetime do not satisfy the Einstein equation.
The fact that taking an average metric and plugging it into the
Einstein equation and plugging the real metric into the equation
and then averaging are not equivalent is sometimes expressed by
saying that because the Einstein tensor $G_{\mu\nu}$ is non-linear
in the metric, we have
$\langle G_{\mu\nu}(g_{\alpha\beta})\rangle \neq G_{\mu\nu}(\langle g_{\alpha\beta}\rangle)$, where $\langle\rangle$ stands for averaging.
It would be more accurate to say that the problem arises because
time evolution and averaging do not commute. (The statement
about averaging the Einstein equation is anyway rather heuristic
because tensors cannot be straightforwardly averaged on curved
manifolds; though see \cite{Zalaletdinov}.)
This is the origin of the fitting problem discussed in \sec{sec:intro}.

The standard assumption is that even after non-linear structures
have started forming, the average evolves according to the FRW
equations if smoothing on the scale of the non-linearity
is performed. However, the equations for the mean expansion
which are actually derived, rather than assumed, do not bear out
this expectation \cite{Buchert:1995, Buchert:1999, Buchert:2001}.
As we will discuss in \sec{sec:Buchert}, inhomogeneities and
anisotropies affect
the behaviour of the scale factor, and neither statistical homogeneity
and isotropy nor the small size of the inhomogeneous and anisotropic
regions (relative to the visual horizon) is a sufficient condition for
recovering the FRW behaviour. 

A different reason why the FRW equations could be invalid even
though the universe is very homogeneous and isotropic on large scales has
also been advanced. It has been suggested that the breakdown
of the approximation of treating the matter as an ideal fluid
involves negative pressure which might explain the acceleration \cite{Schwarz}.
An argument against the treatment of matter as an ideal fluid
goes as follows. Small-scale processes
in the universe are in general not thermodynamically irreversible
but instead produce entropy. Since there is no production of
negative entropy, this entropy generation does not vanish upon
averaging. In contrast, in the approximation of an adiabatically expanding
ideal fluid, the entropy of the universe is constant, and such
small-scale effects are completely absent. In numerical terms,
a single $3\times10^6$ solar mass black hole, such as the one at
the center of our galaxy, has an entropy of $10^{90}$, of the
same order of magnitude or more as the total entropy ascribed to the observable
universe in the ideal fluid picture. Supermassive black holes
are abundant in the universe, so the entropy associated with them
alone completely overwhelms the entropy associated with the adiabatic fluid.
On the other hand, in the average description of an ideal fluid
spacetime it may look as if the entropy was increasing, even though
there is no local entropy production \cite{Hosoya:2004}.
The issue of gravitational entropy is tied up with coarse-graining
and thus backreaction, and is not well understood \cite{Ellis:2005}.
Whether these problems of the ideal fluid description are
important for the expansion rate is not clear.

A related issue is whether one can neglect the influence of
the structure within stabilised regions on the overall expansion,
i.e. whether one can continuously and consistently ``renormalise''
the scale of the stable regions which are treated as particles
of the dust fluid \cite{Padmanabhan:2005}.

We will not study these two issues, and keep to the assumption
that deviations from the dust behaviour are not important for the
overall expansion.

\paragraph{Assumption 3: FRW + perturbations.}

The region of validity of the assumption that inhomogeneities
and anisotropies on a FRW background
evolve according to linear perturbation theory is well-known,
and its breakdown at the end of the linear regime is well understood.
As long as the density contrast $\delta\equiv (\rho-\av{\rho})/\av{\rho}$
of a perturbation is small, it evolves according to linear
perturbation theory. In a spatially flat matter-dominated FRW background,
the growing mode is proportional to the scale
factor, $\delta\propto a$. As $\delta$ becomes of order $\pm1$,
the linear approximation breaks down. An overdensity will collapse
and the density contrast will grow faster than in the linear regime,
whereas an underdensity will grow more slowly than in the linear regime,
asymptotically approaching emptiness.

For intermediate regime perturbations which have not yet gone
non-linear but are nevertheless small enough to be located
inside a non-linear structure, one would expect the evolution
to depend on the environment and not just on the overall average
expansion of the universe.
For example, the evolution of a 20 \mpc\ radius perturbation
would be expected to be different inside a 50 \mpc\ void or a
200 \mpc\ wall. Such perturbations, which are larger than the size
of typical structures and yet fit inside non-linear regions, are by
definition untypical, and one can argue that their effect on the mean
evolution of perturbations is small.

However, it is not obvious that linear perturbation theory
around the average in a space which is highly inhomogeneous
and anisotropic correctly captures the evolution even for
perturbations with wavelength longer than the size of the largest structures.
Regions with large overdensities or underdensities presumably
contribute differently to the evolution of the long-wavelength
modes, and it is not clear that these effects would cancel to give
the same answer as linear perturbation theory around the average.

The separation into background and perturbations is more involved
in a universe with large inhomogeneities and anisotropies than
in the FRW case, and in \sec{sec:Buchert} below we will show that
linear perturbations do not in general satisfy the same perturbation
equation as in FRW models. We will not discuss the issue further, but
in a realistic model of backreaction, the sensitivity of perturbations
to smaller scale inhomogeneities and anisotropies should be looked at
in more detail, particularly when these have a large effect on
the background. For work on perturbations in a backreaction context,
see \cite{Ehlers:1996, Takada:1999, Sicka, Taruya:1999, Tatekawa:2001}.

\subsection{The Buchert equations} \label{sec:Buchert}

\paragraph{The metric and the local equations.}

We assume that the matter content of the universe can be described
as dust, i.e. a pressureless ideal fluid. We further assume that the
dust is irrotational (i.e. the vorticity is zero). Then the metric
can be written in the synchronous gauge \cite{Ehlers:1961, Ellis:1971}
\bea \label{metric}
  \rmd s^2 &=& - \rmd t^2 + {^{(3)}g}_{ij} \rmd x^i\rmd x^j \ ,
\eea

\noindent where $t$ is the proper time measured by observers comoving
with the dust and ${^{(3)}g}_{ij}(t,\bx)$ is the metric on the
hypersurface of constant $t$. The Einstein equation reads
\bea \label{Einstein}
  G_{\mu\nu} &=& 8 \pi G_N T_{\mu\nu} \el
  &=& 8 \pi G_N \rho\, u_{\mu} u_{\nu} \ ,
\eea

\noindent where $G_{\mu\nu}(t,\bx)$ is the Einstein tensor, $G_N$ is
Newton's constant, $T_{\mu\nu}(t,\bx)$ is the energy--momentum tensor,
$\rho(t,\bx)$ is the dust energy density and $u^{\mu}=(1,\bfm{0})$
is the velocity of comoving observers.

We wish to find the equations for average quantities.
Since only scalars can be straightforwardly integrated on a
curved manifold (though see \cite{Zalaletdinov}), we should
project \re{Einstein} to obtain a set of scalar equations. In addition
to $u^{\mu}$ and $g^{\mu\nu}$, we have the covariant derivative
$\nabla^{\mu}$ available. From these we can build three
independent rank two tensors to project with\footnote{Since
$\nabla^{\nu} G_{\mu\nu}$ vanishes identically,
$\nabla^{\mu}\nabla^{\nu}$ does not give anything new compared to
$u^{\mu}\nabla^{\nu}$.}, so the Einstein equation \re{Einstein}
yields the following three exact, local, covariant scalar equations
\cite{Ehlers:1961, Ellis:1971, Raychaudhuri:1955}
\bea
  \label{Rayloc} \dot{\theta} + \frac{1}{3} \theta^2 &=& - 4 \pi G_N \rho - 2 \sigma^2 \\
  \label{Hamloc} \frac{1}{3} \theta^2 &=& 8 \pi G_N \rho - \frac{1}{2} \sR + \sigma^2 \\
  \label{consloc} \rhodot + \theta\rho &=& 0 \ ,
\eea

\noindent where a dot stands for derivative with respect to $t$,
$\theta(t,\bx) = ( \sqrt{^{(3)}g} )^{-1} \pat_t ( \sqrt{^{(3)}g} )$
is the expansion rate of the local volume element,
$\sigma^2(t,\bx)=2\,\sigma^{ij}\sigma_{ij}\geq0$
is the scalar built from the shear tensor $\sigma_{ij}$,
and $\sR(t,\bx)$ is the Ricci scalar of the hypersurface of
constant $t$ (i.e. the spatial curvature).
The acceleration equation \re{Rayloc} is known as the Raychaudhuri
equation, and \re{Hamloc} is the Hamiltonian constraint.

The price for reducing the Einstein equation down to
a set of scalar equations is that the system is not closed:
there are three equations for four independent variables.
Essentially, the propagation of the shear tensor
(or equivalently, of the Ricci tensor on the hypersurface of
constant $t$) does not reduce to a scalar equation.
The integrability condition between \re{Rayloc} and \re{Hamloc} reads
\bea \label{intloc}
  \pat_t ({\sR}) + \frac{2}{3}\theta\ \sR = 2\, \pat_t{\sigma^2} + 4 \theta \sigma^2 \ ,
\eea

\noindent so specifying either the shear or the spatial
curvature fixes the other.

Note that no approximations have been made: the equations
\re{Rayloc}--\re{consloc} are exact for irrotational dust, with
arbitrarily large density variations.

\paragraph{Deriving the Buchert equations.}

We are interested in the evolution of average quantities:
specifically, we want to know how the average expansion rate
behaves. Our discussion follows the original derivation
by Buchert \cite{Buchert:1999}.
When (and only when) the vorticity is zero, the rest spaces
of constant proper time of comoving observers mesh together
to form a family of hypersurfaces which fills spacetime.
The spatial average of a scalar quantity $f$ is then straightforwardly
defined on these hypersurfaces as
\bea \label{av}
  \av{f}(t) \equiv \frac{ \int d^3 x \sqrt{^{(3)}g(t,\bx)} \, f(t,\bx) }{ \int d^3 x \sqrt{^{(3)}g(t,\bx)} } \ ,
\eea

\noindent where the integral is over the hypersurface of constant $t$.
An important property of the averaging \re{av} is that it
does not commute with time evolution,
\bea \label{nc}
  \pat_t\av{f} = \av{\pat_t f} +\av{f\theta} - \av{f} \av{\theta} \ .
\eea

In order to describe the average evolution and compare it to
FRW models, we have to define a scale factor.
The simplest extension of the notion of an overall scale factor to
an inhomogeneous and/or anisotropic spacetime is to define it with
the volume of the hypersurface of constant $t$:
\bea  \label{a}
  a(t) \equiv \left( \frac{ \int d^3 x \sqrt{^{(3)}g(t,\bx)} }{ \int d^3 x \sqrt{^{(3)}g(t_0,\bx)} } \right)^{\frac{1}{3}}  \ ,
\eea

\noindent where the normalisation has been chosen as $a(t_0)=1$
at some time $t_0$ (which we shall take to be today).
In words, $a(t)^3$ is the volume of the hypersurface of constant $t$
(up to the usual multiplicative constant).
We could equivalently define the scale factor with the average of
the volume expansion rate, by $3\adot/a\equiv\av{\theta}$. We will
also use the notation $H\equiv\adot/a$.

By taking the average \re{av} of the scalar equations
\re{Rayloc}--\re{consloc} and commuting the time derivatives
as shown in \re{nc} we obtain the equations satisfied
by the scale factor \re{a}, first derived by Thomas Buchert
in 1999 \cite{Buchert:1999}:
\bea
  \label{Ray} 3 \frac{\addot}{a} &=& - 4 \pi G_N \av{\rho} + \sQ \\
  \label{Ham} 3 \HH &=& 8 \pi G_N \av{\rho} - \frac{1}{2}\av{\sR} - \frac{1}{2}\sQ \\
  \label{cons} && \pat_t \av{\rho} + 3 \H \av{\rho} = 0 \ ,
\eea

\noindent and the integrability condition between the average Raychaudhuri
equation \re{Ray} and the average Hamiltonian constraint \re{Ham},
analogous to \re{intloc}, reads
\bea \label{int}
  \pat_t{\av{\sR}} + 2 \H\av{\sR} = - \dot{\sQ} - 6 \H\sQ \ ,
\eea

\noindent where the backreaction variable $\sQ$ is a new term compared to
the FRW equations, containing the effect of inhomogeneity and anisotropy:
\bea \label{Q}
  \sQ \equiv \frac{2}{3}\left( \av{\theta^2} - \av{\theta}^2 \right) - 2 \av{\sigma^2} \ .
\eea

The Buchert equations \re{Ray}--\re{int} are exact for the averages
when matter consists of irrotational dust. (The Newtonian limit was
derived by Buchert and Ehlers in 1995 \cite{Buchert:1995} and the case
with non-zero pressure by Buchert in 2001 \cite{Buchert:2001}.)
The backreaction variable $\sQ$ consists of two terms:
the variance of the expansion rate and the shear.
Shear is also present in the local equation \re{Rayloc}
and decelerates expansion. In contrast, the variance is only
present in the averaged equations, arising from the the non-commutation
of averaging and taking a time derivative as shown in \re{nc}, and
acts to accelerate the expansion rate.
The presence of this term makes it possible for the average equations
to display behaviour which is qualitatively different from the local behaviour.
In the Newtonian limit, $\sQ$ can be written in terms of the
Minkowski functionals, which are a statistical measure of
the morphological properties of cosmic structure, relating
$\sQ$ directly to structure formation \cite{Buchert:2000}.

\paragraph{Recovering the FRW equations.}

As with the system \re{Rayloc}--\re{intloc}, there are only three
independent equations in \re{Ray}--\re{int} for the four independent
functions $a, \av{\rho}, \sQ$ and $\av{\sR}$. Physically this means that
different inhomogeneous and/or anisotropic spacetimes can evolve
differently even if they have the same average initial conditions.
While the equations \re{Ray}--\re{int} cannot be
solved, they can be used to check whether a given scale factor can
result from backreaction \cite{Rasanen:2005}.

Specifying one more function (or relation between the four functions)
leads to a soluble system. In particular, in the limit when the shear
and the variance of the expansion rate are small compared to the
contribution of the energy density, the Buchert equations
reduce to the FRW equations. The integrability condition \re{int}
between \re{Ray} and \re{Ham} then leads to the standard behaviour
$\av{\sR}\propto a^{-2}$.
(Since the integrability condition ties the evolution of spatial
curvature to the backreaction variable $\sQ$, the average spatial
curvature evolves like $\av{\sR}\propto a^{-2}$ {\it only} in the
FRW limit or in the special case when $\sQ\propto a^{-6}$, in
agreement with the discussion of assumption 1 in \sec{sec:FRW}.)
Conversely, if the shear or the variance of the expansion rate
are large compared to the contribution of the energy density
in a large fraction of space, the FRW equations are not a good
approximation for the average behaviour.
This is the way to derive the FRW equations and quantify
their domain of validity.

The average behaviour also reduces to the FRW equations
when the shear and the variance of the expansion rate cancel,
even if they are not small. (For examples of such solutions,
see \cite{Kasai}.) In the Newtonian limit, this happens
generically for periodic boundary conditions \cite{Buchert:1995}
and for spherically symmetric spaces \cite{Sicka}.
(With Minkowski functionals, the latter result can be understood
to follow from the property that the backreaction variable
$\sQ$ in the Newtonian limit measures the deviation of
morphology from that of a ball \cite{Buchert:2000}.)
This cancellation is also present in relativistic perturbation theory
when expanding to second order, as shown in \cite{Rasanen:2003}
and recently rederived in \cite{Kasai:2006}.
This is not evidence towards a theorem that backreaction would
not in general affect the acceleration, as claimed in \cite{Kasai:2006}.
Instead, it is related to the vanishing of backreaction for periodic
boundary conditions in the Newtonian limit.
It was correctly identified in \cite{Notari:2005} that in the
perturbative expansion the Newtonian terms are those with the 
largest number of spatial gradients, as they are accompanied by the
largest number of powers of the speed of light. 
Therefore, the terms with the highest number of spatial gradients
at each order in perturbation theory vanish for periodic boundary
conditions (which are implicit in the use of Fourier decomposition).
Since the highest number of gradients at order $N$ in perturbation
theory is $2 N$, the second order term with four gradients vanishes
upon averaging. However, at fourth order there is no reason for the
term with six derivatives to vanish, and this term becomes large
after non-linear structures start forming, so the perturbation
expansion is expected to break down.

An important point is that the expansion described by the Buchert
equations \re{Ray}--\re{int} does {\it not} reduce to the FRW
behaviour simply when the spatial size of the inhomogeneities and
anisotropies is small. If the shear or the variance of the expansion rate
is comparable to the contribution of the energy density in a sizeable
fraction of space, the behaviour will deviate from the FRW case
(barring the sort of cancellation discussed above),
regardless of the size of the individual inhomogeneous and/or
anisotropic regions.
The variance of the expansion rate is non-negative, and only
vanishes if the expansion rate is completely homogeneous: the
contributions from different inhomogeneous regions cannot
cancel to zero. The same is true for the shear. (Of course, the
shear and variance from different regions could cancel each other.)
The Buchert equations \re{Ray}--\re{int} make explicit
and quantify the statement made in \sec{sec:FRW} that
large-scale statistical homogeneity and isotropy does not
guarantee that the dynamics of the scale factor follows the FRW equations.

\paragraph{Acceleration without acceleration.}

The average Raychaudhuri equation \re{Ray} shows
that if the variance of the expansion rate is large enough
compared to the shear and the energy density, the average
expansion accelerates, even though the
local expansion rate decelerates at every point
according to the local Raychaudhuri equation \re{Rayloc}.
The possibility of acceleration is a property of the averaged
system which is not present in the local behaviour, and is
due to the non-commutation of averaging and time evolution.

Since $\sQ$ contributes positively to the acceleration
\re{Ray}, but negatively to the Hubble rate \re{Ham}, it
might seem that negative curvature is needed to balance
the negative contribution of $\sQ$ to the Hubble rate, as
claimed in \cite{Rasanen:2005}.
In fact, it is possible to have acceleration even when the spatial
curvature is positive, as all that is needed is that the sum
$-\av{\sR}-\sQ$ increases (i.e. becomes less negative).
This just means that $\av{\sR}$ has to decrease
(i.e. $-\av{\sR}$ has to increase) faster than $\sQ$ is growing.
Acceleration necessarily involves (non-FRW) spatial curvature,
as \re{int} shows: if $\av{\sR}\propto a^{-2}$,
we have $\sQ\propto a^{-6}$, and there is no acceleration.
This feature demonstrates the difference between having a
description in terms of a scale factor and the FRW metric
being valid, discussed in the context of assumption 1 in \sec{sec:FRW}.

One consequence of the fact that local expansion can only decelerate
is that once a shell of matter has started collapsing, it cannot
turn around and stabilise. It follows that
the formation of stable structures involves vorticity and/or breakdown
of the dust approximation \cite{DelPopolo:2001, Buchert:2005a}.
The contribution of vorticity to the local acceleration \re{Rayloc}
(which we did not include) is always positive, so it cannot disappear
upon averaging, as the contribution
from each region has the same sign. This is also
the case for shear, and the two can cancel each other.
Indeed, in stabilised regions, the positive contribution of vorticity
has to exactly balance the negative contributions of the shear and
the energy density to produce net zero acceleration.
One would thus naively expect our approximation of neglecting
vorticity to lead to a lower bound on the acceleration.
However, vorticity would also complicate the averaging proceduce,
as discussed in \sec{sec:intro}, so the issue should be carefully studied.

We will discuss the physical meaning of the average acceleration
and the relation to structure formation in \sec{sec:acc},
but let us first complete the overview of the FRW assumptions
by looking at assumption 3 concerning the behaviour of small
perturbations in an inhomogeneous and/or anisotropic universe.

\paragraph{The evolution of perturbations.}

We will briefly consider perturbation theory to see what are
the corrections to the FRW picture.
For linearly perturbed FRW models, one writes
the equations of motion as the sum of the background part
plus a small perturbation. Discarding terms beyond first order
in the perturbation and taking the spatial average
leads to the equations for the FRW background
(since the average of the perturbation is taken to vanish).
Deducting these from the perturbed equations then gives the
evolution equation for the perturbations.

When backreaction is important, the same procedure
does not work, as the difference between the local equations
\re{Rayloc}--\re{consloc} and the averages \re{Ray}--\re{cons}
is large by definition. We will do the closest thing,
which is to separate the local terms into a 'large' part
which determines the average behaviour and a 'small' part
which does not contribute to the averages.
We write $\theta(t,\bx)=\theta_0(t,\bx)+\Delta\theta(t,\bx)$,
and assume that $\abs{\theta_0}\gg\abs{\Delta\theta}$.
We similarly split $\rho(t,\bx)=\rho_0(t,\bx)+\Delta\rho(t,\bx)$,
$\sigma^2(t,\bx)=\sigma^2_0(t,\bx)+\Delta(\sigma^2)(t,\bx)$
and $\sR(t,\bx)=\sR_0(t,\bx)+\Delta(\sR)(t,\bx)$.
The assumption that the 'large' part is responsible for the
average behaviour, $\av{\theta}=\av{\theta_0}$, implies that
$\av{\Delta\theta}=0$, and similarly for the other quantities.
We then insert this ansatz into the local equations
\re{Rayloc}--\re{consloc} and keep only terms up to linear order
in the 'small' quantities. Unlike in the FRW case, there is
no rigorous way to separate the perturbation from the background,
as both have spatial dependence. A similar issue arises in
deriving the perturbation equations for 'tilted' cosmological models,
i.e. models where the fluid velocity is not everywhere normal
to the hypersurface of constant proper time \cite{vanElst:1998}.
(If we had  non-zero vorticity, the model would necessarily be tilted.)
If we simply assume that the equations
for the 'large' and 'small' parts decouple, then the 'large' parts
satisfy the local equations \re{Rayloc}--\re{consloc} and
their average gives the Buchert equations \re{Ray}--\re{cons}.
For the 'small' terms, we then obtain the following evolution equation
\bea \label{pert1}
  \deltaddot + \frac{2}{3} \theta_0 \deltadot - 4 \pi G_N \rho_0 \delta = 2 \Delta(\sigma^2) \ ,
\eea

\noindent where $\delta\equiv\Delta\rho/\rho_0$, along with
the consistency condition $\av{\theta_0\Delta\theta}=0$.
In addition to the shear source term on the right-hand side,
\re{pert1} differs from its FRW counterpart in that
$\theta_0, \rho_0$ are position-dependent.

It is possible to obtain evolution equations where the average
expansion rate and energy density appear instead of the local quantities.
Let us assume that $\delta$ is separable, $\delta(t,\bx)=D(t)h(\bx)$;
in the linearly perturbed FRW case, this would correspond to
considering pure growing or decaying modes. Dividing \re{pert1}
by $h(\bx)$ and averaging, we have
\bea \label{pert2}
  \ddot{D} + 2 \H \dot{D} - 4 \pi G_N \av{\rho} D = 2 \left\langle \frac{\Delta(\sigma^2)}{h} \right\rangle \ .
\eea

\noindent The evolution equation \re{pert2} agrees with the linearly
perturbed FRW case\footnote{For a given $a(t)$; the evolution of
$a$ (and thus the evolution of $D$) as a function of time will
in general be different, since it is governed by the Buchert
equations and not by the FRW equations.}, apart from the shear source
term (note that $\av{\rho}\propto a^{-3}$ by \re{cons}).
In the spatially flat FRW limit, we recover the standard behaviour
$D\propto a$ for the growing mode. Let us write $D(t)=a(t)d(t)$
in order to analyse the deviation from this limiting case. Using
equations \re{Ray} and \re{Ham} we can write \re{pert2} as
\bea \label{pert3}
  \frac{\ddot{d}}{d} + 4 \H \frac{\dot{d}}{d} = \frac{1}{3} \av{\sR} + 2 \frac{1}{a d} \left\langle \frac{\Delta(\sigma^2)}{h} \right\rangle \ .
\eea

If the expansion accelerates, the spatial curvature will
become more negative, tending to decrease $d$,
just like acceleration and negative spatial curvature 
act against the growth of perturbations in FRW models.
The shear term can have either sign: writing
$\sigma^{ij}=\sigma^{ij}_0+\Delta\sigma^{ij}$ for the shear tensor,
we have $\Delta(\sigma^2)=4\,\sigma^{ij}_0\Delta\sigma_{ij}$.

Just as the equations for the averages are not closed, neither
are the perturbation equations. The perturbation equations
require more information about the inhomogeneities and anisotropies
(specifically about the shear perturbation) in addition to that
needed for closure of the average equations.
Once this information is supplied, the
backreaction equations for the averages and the perturbations
are analogous to modified gravity: there is no new energy component,
but the relation between the average energy density and the average
expansion rate, as well as between the averages and the perturbations,
is different from the FRW case.
For the averages, any model of modified gravity can of course be
formally written as a general relativistic model with extra sources,
and vice versa. Likewise, in the case of backreaction one can write
the average behaviour in terms of a scalar field, the ``morphon''
\cite{Buchert:2006}. It would be interesting to see how far the
analogy can be extended to the perturbations.

We will not discuss perturbation theory further. In a realistic model
of backreaction the issue should be studied in detail, in order
to be able to compare with observations of CMB and LSS. For work
on perturbations in a backreaction context, see
\cite{Ehlers:1996, Takada:1999, Sicka, Taruya:1999, Tatekawa:2001}.

\section{Acceleration from collapse} \label{sec:acc}

\subsection{A toy model for backreaction} \label{sec:toy}

\paragraph{Breakdown of the FRW approximation.}

The proposal \cite{Rasanen:2003} that structure formation leads
to accelerating expansion implies in the context of the
Buchert equations \re{Ray}, \re{Q} that the relative variance
of the expansion rate is of order one. This behaviour has been
numerically demonstrated in specific examples of LTB models
\cite{Chuang:2005, Kai:2006, Paranjape:2006}.
However, the physical meaning of having average acceleration
even though the local expansion decelerates everywhere has not been clear
\cite{Ishibashi:2005} (see also \cite{Kolb:2005c}).

We will clarify the physical meaning of the acceleration with a
simple toy model, and show that it is not unreasonable for
structure formation to involve a variance in the expansion rate
that is large enough to produce acceleration.
A shorter treatment was presented in the essay \cite{Rasanen:2006},
and the role of collapsing regions has also been discussed in \cite{Kai:2006}.

As mentioned in \sec{sec:FRW} when discussing assumption 3,
the breakdown of perturbation theory at $\delta\sim\pm1$
is well understood. The simplest model used to describe
the non-linear evolution is the spherical collapse model
for overdense regions (see e.g. \cite{Padmanabhan:1993, Liddle:2000}),
and the equivalent for underdense regions (the appendix of
\cite{Sheth:2003} has a useful summary of both cases).
The model consists of a spherically symmetric density perturbation
embedded in a FRW universe (surrounded by an empty region to
make the mean density agree with the background), studied
in the Newtonian limit. The perturbation behaves on average like
an independent FRW universe with positive or negative spatial
curvature in the case of an overdense or underdense region,
respectively. In terms of the Buchert equations \re{Ray}--\re{int},
this comes about because the backreaction variable
$\sQ$ vanishes for spherical symmetry in the Newtonian
limit \cite{Sicka}. (For some work on the general relativistic
case, see \cite{Rasanen:2004, Paranjape:2006}.)

According to the spherical collapse model, the expansion of an
overdense domain slows down with respect to the background until
the structure stops expanding, turns around and starts collapsing.
The spherical collapse model indicates that matter collapses
to a singularity in a finite time, whereas in practice structures
stabilise at a finite radius, usually taken to be half the
radius at turnaround. The spherical collapse model is not a good
description of the final stages of collapse, as
departures from spherical symmetry are important for real
structure formation, and collapse amplifies asymmetries.
There are more accurate and realistic treatments of collapse,
see e.g. \cite{Taruya:1999, Kerscher:2000, DelPopolo:2001, Engineer:1998},
but as we are mostly interested in qualitative features, 
the spherical collapse model will be an adequate description
of non-linear density perturbations.

What is not generally appreciated is that just as linear perturbation
theory around the FRW universe breaks down as perturbations become
non-linear, the FRW universe itself breaks down as a description
of the average behaviour when the non-linear perturbations occupy
a sizeable fraction of space, as discussed in \sec{sec:Buchert}.
We want to look at this breakdown of FRW equations with a simple
model of structure formation, in analogy with the description
of the breakdown of perturbation theory in the spherical collapse model.

\paragraph{Two-region toy model of structure formation.}

In the real universe, structure formation consists
of small overdensities and underdensities
developing into stable structures with fixed density and voids
which are constantly becoming emptier, respectively.
We will consider the simplest possible toy model of structure
formation, with two separate spherically symmetric dust regions,
one overdense and one underdense. We will consider the Newtonian limit,
so that the regions evolve according to the spherical collapse model.
The regions are taken to be disjoint, and we ignore their embedding
into the whole space. We are then essentially just
comparing two FRW universes. The same kind of toy model was
qualitatively discussed in \cite{Ishibashi:2005}; we will take a
more detailed look and add an understanding of the physics behind
the equations.

We denote the scale factors of the regions by $a_1, a_2$ and
the corresponding Hubble parameters by $H_1, H_2$, where
region 1 is underdense and region 2 is overdense, so
$H_1>H_2$. Since the regions are disjoint, the total
volume is simply the sum of the volumes $a_1^3$ and $a_2^3$, and the
overall scale factor is $a=(a_1^3+a_2^3)^{1/3}$. The overall Hubble
and deceleration parameters are (they can be computed from \re{Ray},
\re{Ham}, \re{Q} or directly from the definition of $a$)
\bea
  \label{Hex} \!\!\!\!\!\!\!\!\!\!\!\!\!\!\!\!\!\!\!
H &=& \frac{ a_1^3 }{ a_1^3 + a_2^3 } H_1 + \frac{ a_2^3 }{ a_1^3 + a_2^3 } H_2 \equiv H_1 \left( 1 - v + v h \right) \\
  \label{qex} \!\!\!\!\!\!\!\!\!\!\!\!\!\!\!\!\!\!\!
q &\equiv& - \frac{1}{H^2} \frac{\addot}{a} = q_1 \frac{ 1-v }{ (1-v+h v)^2 } + q_2 \frac{ v h^2 }{ (1-v+h v)^2 } - 2 \frac{ v (1-v) (1-h)^2  }{ (1-v+h v)^2 } \ ,
\eea

\noindent where $q_1, q_2$ are the deceleration parameters of
regions 1 and 2, and we have denoted the fraction of space in
the overdense region by $v\equiv a_2^3/(a_1^3+a_2^3)$ and
the relative expansion rate of the two regions by $h\equiv H_2/H_1$.

The Hubble rate \re{Hex} is simply the volume-weighted average
of the Hubble rates in regions 1 and 2. Not so for the deceleration
parameter: in addition to the first two terms in \re{qex},
there is a third term related to the variance of the expansion rate,
corresponding to the backreaction variable $\sQ$ in the average
Raychaudhuri equation \re{Ray}. While $q_1, q_2$ are positive or
zero, the last term is always negative, corresponding to the fact
that $\sQ$ is positive.

For simplicity, we take the underdense region to be completely empty,
so $a_1\propto t$. Region 2 behaves like a closed FRW universe, with
$a_2\propto 1-\cos\phi$, $t\propto \phi-\sin\phi$, where
the parameter $\phi$ is called the development angle. The overdense
region starts expanding from the big bang singularity at $\phi=0$ and
slows down until it turns around at $\phi=\pi$ and starts collapsing,
finally shrinking to zero size and infinite density at $\phi=2\pi$.
In practice, overdense regions stabilise at fixed size
and density. In the spherical collapse model, this is
often implemented by hand at $\phi=3\pi/2$, when the radius
of the structure is half the radius at turnaround. We will
therefore follow the evolution only until $\phi=3\pi/2$.

There is one free parameter in this toy model, the relative size of the
two regions at some given time. Taking this time to be at turnaround
at $\phi=\pi$ and denoting the fractions of space in regions 1 and 2
at that moment by $f_1=1-f_2, f_2$, we have
\bea \label{vandh}
  v &=& \frac{ f_2 \pi^3 (1-\cos\phi)^3 }{ 8 (1-f_2) (\phi-\sin\phi)^3 + f_2 \pi^3 (1-\cos\phi)^3 } \el
  h &=& \frac{ \sin\phi (\phi-\sin\phi) }{ (1-\cos\phi)^2 } \ .
\eea

\noindent Inserting \re{vandh} into \re{qex}, it is easy to establish that
the acceleration can be positive.

\begin{figure}
\hfill
\begin{minipage}[t]{5.1cm} 
\scalebox{1.2}{\includegraphics[angle=0, clip=true, trim=0cm 2cm 0cm 0cm, width=\textwidth]{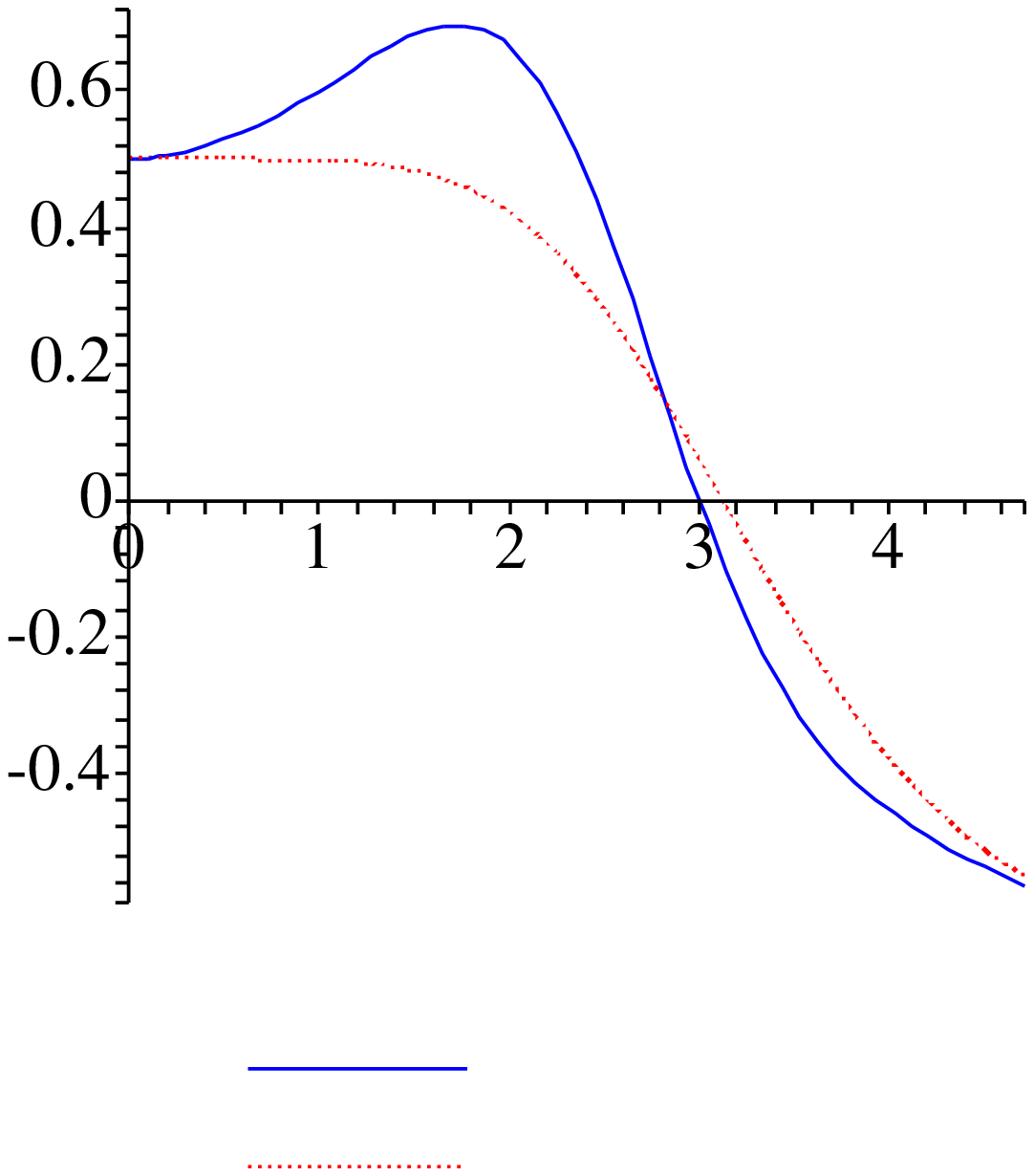}}
\begin{center} {\bf (a)} \end{center}
\end{minipage}
\hfill
\begin{minipage}[t]{5.1cm}
\scalebox{1.2}{\includegraphics[angle=0, clip=true, trim=0cm 2cm 0cm 0cm, width=\textwidth]{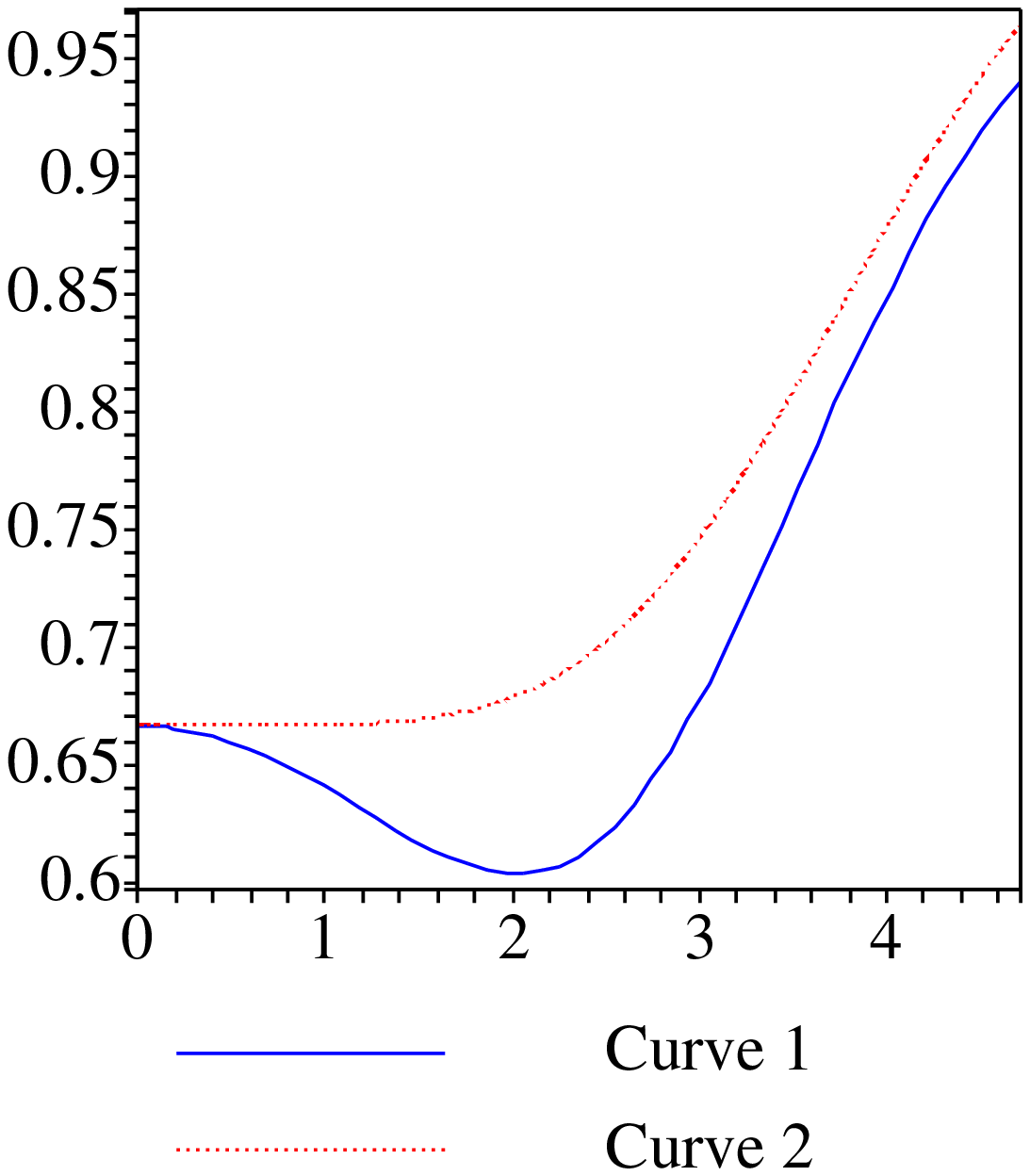}}
\begin{center} {\bf (b)} \end{center}
\end{minipage}
\hfill
\begin{minipage}[t]{5.1cm}
\scalebox{1.3}{\includegraphics[angle=0, clip=true, trim=0cm 3.5cm 0cm 0cm, width=\textwidth]{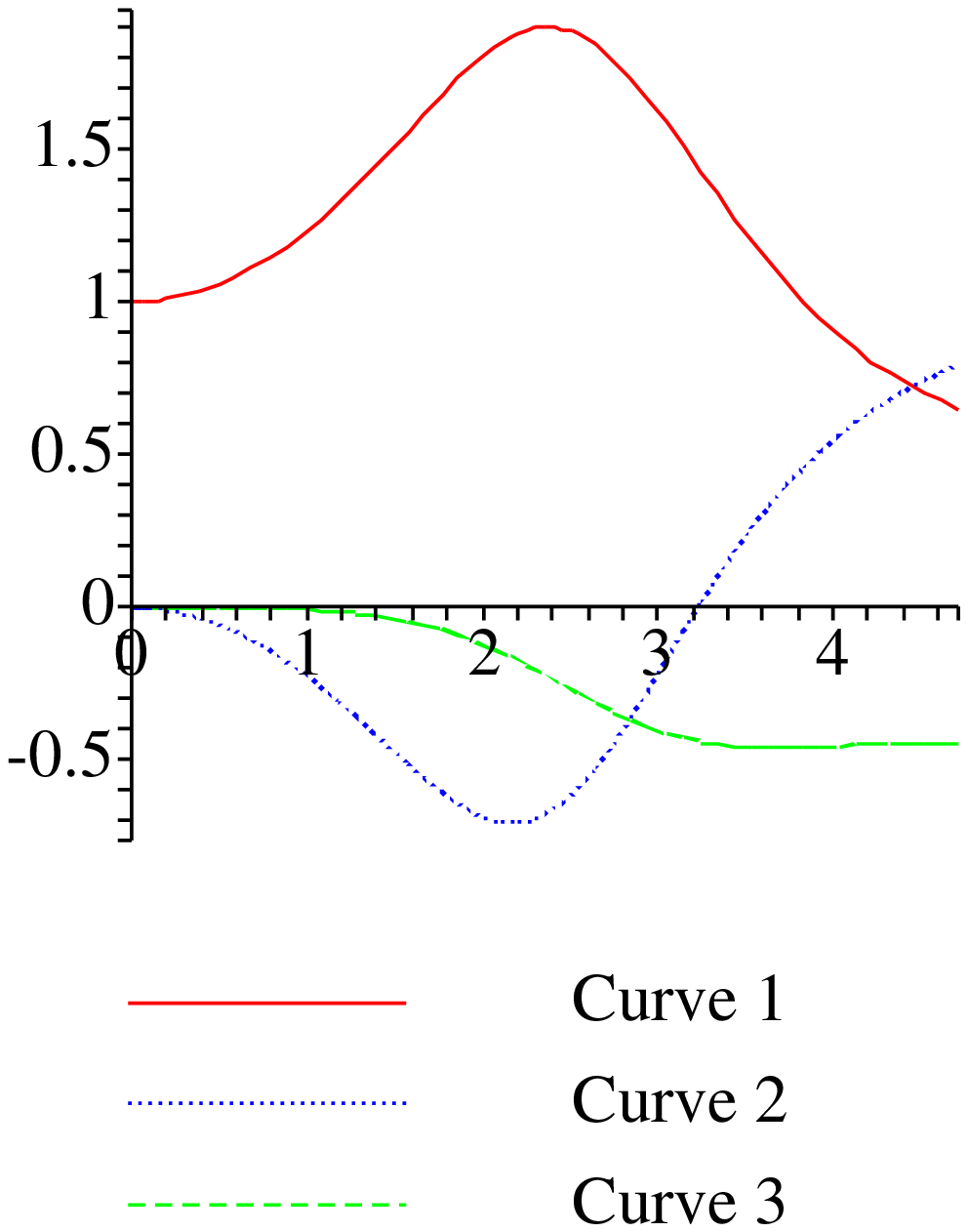}}
\begin{center} {\bf (c)} \end{center}
\end{minipage}
\hfill
\caption{The evolution of the toy model as a function of the
development angle $\phi$.
(a): The deceleration parameter $q$ in the toy model (blue, solid)
and in the \LCDM model (red, dash-dot).
(b): The Hubble parameter multiplied by time, $H t$, in the toy model
(blue, solid) and in the \LCDM model (red, dash-dot).
(c): The density parameters of matter, spatial curvature
and the backreaction variable $\sQ$ in the toy model.
Red (solid) is $\Om$, blue (dash-dot) is $\OR$ and green (dash) is $\OQ$.}
\label{fig:toy}
\end{figure}

In \fig{fig:toy} (a) and (b), we have plotted the deceleration parameter
$q$ and the Hubble rate multiplied by $t$, as dimensionless
measures of the acceleration and the expansion rate, respectively.
In addition to the toy model, we show the behaviour in the \LCDM model,
just to make the qualitative features of backreaction-driven acceleration
easier to grasp by comparison; the toy model is not meant
to be taken seriously in a quantitative sense.
We have chosen the value of the free parameter to be $f_2=0.3$,
so that the value of $q$ at $\phi=3\pi/2$ in the toy model
approximately equals the value in the \LCDM model when
$\Om=0.3, \Omega_{\Lambda}=0.7$.
In \fig{fig:toy} (c) we show the density parameters
for matter, spatial curvature and the backreaction variable $\sQ$
(defined in \sec{sec:obs} after \re{q}, \re{Omegas}) in the toy model.
Note that negative $\OR$ corresponds to positive spatial curvature,
and vice versa.

Figure \ref{fig:toy} (a) shows that the expansion accelerates,
particularly after region 2 starts collapsing at $\phi=\pi$.
It may seem paradoxical that gravitational collapse
induces acceleration. However, the explanation is simple.
Initially, the expansion is similar to the Einstein--de Sitter case,
with $q=1/2, H t=2/3$. The contribution of the overdense region
$v H_2$ gradually slows down the average expansion rate $H$.
The relative volume $v$ occupied by the overdense region decreases
monotonously, and eventually it becomes so small that the contribution
of the underdense region begins to dominate and the expansion rate,
as measured by $H t$, grows. The related change from positive to
negative spatial curvature is clearly seen in \fig{fig:toy} (c).
This effect is particularly pronounced and easy to understand
once the overdense region has started collapsing: then the 
Hubble rate $H_2$ is actually negative, and its contribution
shrinks rapidly as $a_2$ contracts, so the average expansion rate rises.
(The fraction of volume in the overdense region at $\phi=3\pi/2$
is only $v\approx0.01$.)
The increase in the absolute value of the collapse rate cannot
compensate for the decrease in volume, as the FRW Hubble law shows:
$a_1^3 H_1 = a_1^3 \sqrt{8\pi G_N\rho_{10} a_1^{-3}/3 - K_1 a_1^{-2}}$.
This also means that the results are not sensitive to the
diverging behaviour in the final stages. The contribution
of the collapsing region decreases as it approaches the
singularity, and the mean quantities would remain finite
even if we followed the system to the end of the collapse.

Note that we have first averaged over the internal behaviour of
each region, and only then taken the average over the two regions.
This neglects the variance within each region, thus underestimating the
total variance by the volume-weighted sum of the individual variances.
Since the regions are spherically symmetric, this contribution
to the backreaction variable $\sQ$ (defined in \re{Q}) is exactly
cancelled by the shear contribution, so the two-step method does
give the right $\sQ$.

\paragraph{General lessons.}

The reason that the average expansion can accelerate even though
the local expansion decelerates everywhere is that the growth of
the relative volume occupied by the faster expanding regions
contributes to the average acceleration.
This makes the physical content of the Buchert equation
\re{Ray} clear: the larger is the variance of the expansion rate,
the wider is the difference between the fastest and the slowest
expanding regions, so the more rapidly the relative volume of
the fastest region can grow, and the stronger is the acceleration
(assuming that the shear is not too large).

It is also transparent why acceleration from backreaction involves
decreasing spatial curvature, as indicated by \re{Ray}, \re{Ham}, \re{Q}
and shown by \fig{fig:toy} (c):
the more negatively curved a region is, the faster it expands,
and the larger its relative volume will become. Therefore the most
negatively curved regions will come to dominate the average curvature.
For underdense FRW regions, $H t$ is above $2/3$ and approaches unity
from below. So, the decrease of spatial curvature is expected to be
accompanied by $H t$ approaching $1$, as seen in \fig{fig:toy} (b).

While the rise of $H t$ beyond the Einstein--de Sitter
value of $2/3$ is due to the underdense region,
it is the overdense region which is essential for
acceleration, i.e. the fall of $q$. (It is possible, but harder,
to have acceleration without a collapsing region.)
Unlike in the \LCDM model, these two effects are distinct.
For example, if we replaced the empty void obeying $a_1\propto t$ with
the Einstein--de Sitter universe for which $a_1\propto t^{2/3}$, there
can still be acceleration because of the collapse, even though
$H t$ is at most $2/3$. The important thing is that $H t$ first
slows down, so that it can later rise rapidly due to the collapsing region.

In fact, a more slowly expanding underdense region can even lead
to stronger acceleration. For example, if instead of $a_1\propto t$
we were to take $a_1\propto t^{4/5}$, which is arguably closer to the
behaviour of real voids \cite{voids}, $q$ would be more negative than
in the case when region 2 is empty or an Einstein--de Sitter universe.
The reason is that there are two competing effects: if the expansion
of the faster expanding region is too slow, the contrast with the
overdense region won't be strong enough to give a large variance,
while if it's too rapid, the faster expanding region will dominate
the Hubble rate before the overdense region can have any impact.

The fact that gravity is attractive (for matter satisfying the
strong energy condition) implies that the local
expansion rate is bounded from above, as can be shown by integrating
the Raychaudhuri equation \re{Ray} as an inequality. 
The same holds for the mean expansion rate, resulting in the bound
$H t<1$ for acceleration driven by backreaction \cite{Rasanen:2005}.
In contrast, the collapse rate is not
bounded from below, so collapsing regions can lead to
arbitrarily rapid change in the expansion rate, and $q$
can become arbitrarily negative. In fact, $q$ can diverge to
negative infinity in a finite time.
In the two-region toy model, for sufficiently large $f_2$ the negative
contribution $v H_2$ will at some point equal the positive contribution
$(1-v) H_1$, so $H$ passes zero from above and $q$ diverges to positive
infinity. When $H$ later passes zero from below on its way back to
positive values, $q$ will diverge to negative infinity.

Such behaviour is in contrast to FRW models, where $H t$ can grow
without limit, but $q<-1$ requires violating the null energy
condition (or the modified gravity equivalent).
Acceleration with $q<-1$ is consistent with observations,
and at present there is no statistically significant evidence
either for it or against it \cite{Shapiro:2005, Elgaroy:2006, trans}.

\subsection{Backreaction in the real universe} \label{sec:real}

\paragraph{Start of the acceleration.}

The toy model studied above shows that gravitational collapse is
intimately associated with accelerating expansion,
and makes transparent the physical content of
the Buchert equations \re{Ray}--\re{int}.
Obviously, the real universe does not consist of two disjoint
spherically symmetric regions. Let us clarify which features
of the toy model are expected to be relevant for the real universe.
The density distribution of the universe is characterised
by a hierarchy of perturbations nested within perturbations.
In the past the perturbations were small on
all scales, so the universe was well described by linear
perturbation theory around the FRW universe.
(And it has been shown that a solution of the linearised
equations indeed corresponds to a nearly FRW solution
of the the full equations \cite{Death:1976}.)
In the simplest models of inflation, the
primordial spectrum of density perturbations is adiabatic
and nearly scale-invariant (at horizon entry), and
for typical models of supersymmetric weakly interacting
dark matter there is a cut-off in the power spectrum at small
scales due to collisional damping and free-streaming \cite{SUSYCDM}.

The density perturbations grow logarithmically during the
radiation-dominated era and linearly during the matter-dominated era.
For typical supersymmetric dark matter, the first generation of
perturbations becomes non-linear and forms bound structures
around a redshift of 40-80 \cite{SUSYCDM}.
For other viable dark matter candidates such as axions \cite{axion},
light dark matter \cite{LDM} or right-handed neutrinos \cite{nuMSM},
structure formation may begin at a different time due to differences
in the details of the perturbation spectrum, but the qualitative
picture is expected to remain the same. For a comprehensive
analysis of dark matter candidates, see \cite{Boehm}.

The structures which collapse and stabilise (or form voids) are
part of larger scale perturbations which are in turn
undergoing the process of slowing down and collapsing (or speeding up and
becoming emptier), and so on, with ever larger non-linear overdense
structures and underdense voids forming over time.
Since total mass is conserved, the formation of high-density regions
is accompanied by the formation of larger underdense regions.

One possibility is that the gravitational collapse associated with
the formation of the first generation of structures would lead to
acceleration. Since acceleration smoothens inhomogeneities and anisotropies
and impedes structure formation, backreaction-driven acceleration
will eventually end. This is implicit in the limit $H t<1$, which
rules out eternal acceleration \cite{Rasanen:2005}. 
However, in the case of the real universe the physical reason for
the end of the acceleration is not as clear as in the two-region
toy model, though according to the perturbation equation \re{pert3}
the negative spatial curvature will work against structure formation.
After acceleration ends, inhomogeneities and anisotropies can
become important again, as perturbations are nested inside each other,
with new modes constantly entering the horizon. This could lead
to oscillations between deceleration and acceleration. Such oscillations
are not ruled out observationally, and might be detectable in the
next generation of experiments \cite{osc}.
Oscillation between acceleration and deceleration is a
possible solution to  the coincidence problem, as one could
observe acceleration in the recent past at all times after
the start of structure formation.

The question of whether acceleration does start with the
collapse of the first generation of structures and then
undergo oscillations should be addressed in a realistic model.
At first sight, such early acceleration seems unlikely because
backreaction involves dynamical spatial curvature, as we have discussed.
It seems likely that significant spatial curvature from
redshift $\approx$ 40-80 onwards would have been apparent
in observations, particularly in the CMB \cite{Spergel:2006}.
However, precisely because the spatial curvature does not behave
in the same way as in FRW models, the CMB bounds cannot be
straighthforwardly applied; see \sec{sec:obs} for discussion of
spatial curvature and observations.

Oscillations aside, backreaction offers another way of addressing
the coincidence problem. In addition to the time when the first
generation of objects forms, structure formation involves a second
preferred time, near the era $\sim$ 10 billion years when acceleration
has been observed.

The process of hierarchical structure formation is not entirely
self-similar even when the primordial spectrum of perturbations
is scale-invariant. In addition to the cut-off scale due to
collisional damping and free-streaming, there is at least
one other scale present, related to the
transition from radiation domination to matter domination (the physics
of dark matter or inflation can of course involve further scales).
The scale $k_{eq}$ corresponding to the wavenumber of the mode which
entered the horizon at matter-radiation equality is imprinted on the
perturbation spectrum because metric perturbations which entered the
horizon during the radiation-dominated era ($k>k_{eq}$) are damped
relative to those which enter during matter domination ($k<k_{eq}$).
In the real universe, the matter-radiation equality scale is
$k_{eq}^{-1}\approx$ 60-160 \mpc\ (for $0.15\lesssim\Omn\lesssim0.35$
\cite{Peebles:2004} and the updated value $H_0=62.3\pm5.2$ km/s/Mpc 
\cite{Sandage:2006}).
Therefore the amplitude of the metric perturbations grows with increasing
wavelength, asymptotically approaching the value $A\approx 10^{-5}$
set by inflation (or some other process in the primordial universe).

There is a factor $k^2/(a H)^2$ in going from the metric
perturbations to density perturbations, and the suppression for
small wavelengths is approximately $1/k^2$ up to a logarithm,
so there is only a logarithmic difference in the amplitude of
the density perturbations at small wavelengths, as mentioned above.
As a result, the
size of structures which are just starting to collapse relative
to the Hubble radius, $R_{nl}/(a H)^{-1}$, grows rapidly
at first and rises monotonically, slowing down as structure
formation proceeds and saturating at the value
$\approx\sqrt{A}$ once all perturbations which entered the horizon
during the radiation-dominated era have collapsed.
(Here $R_{nl}(t)$ is the scale at which the mean square of the density
contrast at time $t$ is unity, $\av{\delta^2}(R_{nl}(t),t)=1$.)
If the universe followed the Einstein--de Sitter behaviour, the evolution
would enter the saturation regime around 10-100 billion years.

Once structure formation has started, part of the universe is
in a constant state of collapse, and part is always becoming more
empty. Based on the Buchert equations \re{Ray}--\re{int} and the
analysis of the two-region toy model, we would expect the
backreaction in the real universe to be strongest when the
collapsing objects contract from having occupied the maximum
fraction of volume.
Neglecting evolution in number density, this naively occurs after
the objects have reached their maximum size. The issue is not
entirely clear, however, and it could be that the important factor
is the slowing down of the growth of the collapsing regions as their
size relative to the horizon becomes practically saturated.
At any rate, regarding the coincidence problem,
it is encouraging that these simple arguments lead to a time which
is in rough agreement with the era when acceleration has been observed.

There is another reason to think that acceleration will not occur
with the formation of the first bound objects, but later in the
history of the universe. If we set up the two-region toy model 
discussed in \sec{sec:toy} so that the initial volumes
and absolute magnitudes of the density perturbations in the
overdense and underdense region are equal (which is arguably closer
to the real situation at early times), the expansion goes smoothly
from the Einstein--de Sitter behaviour ($a\propto t^{2/3}$) to
the empty universe behaviour ($a\propto t$). There is no slowdown
period and no acceleration, as the variance of the
expansion rate is too small. The calculation that was done
in the toy model is more representative of structure formation
in the late-time universe, when voids already occupy a large fraction
of the volume.

So, it is plausible that acceleration only occurs once
structure formation takes place in an environment dominated
by voids. In view of this it is, again, encouraging that
simulations and analytical work show that voids grow to
fill practically the whole space, with the void distribution
dominated by voids of a characteristic size, which grows as
larger structures become non-linear and smaller voids merge
\cite{Shandarin, Sheth:2003, Dubinski:1993, Weygaert:2004}.
Not all analyses of simulations agree on the distribution
of void sizes, evolution of the fraction of space occupied by
voids or the fraction of space in voids today.
In \cite{Colberg:2004} the growth of the void volume fraction is clear,
but the value today is only $\approx 0.6$, whereas
the authors of \cite{Furlanetto:2005}
find that the fraction reaches $\approx 1$ at present day. 
Some of the simulations use the \LCDM model, where the present day
is singled out by construction, so care must be taken in applying the
results to a dust-dominated universe. Nevertheless, given that the
contribution of the cosmological constant in small at high redshifts,
one would expect the feature that voids do not dominate early on
to remain valid, while the behaviour at low redshifts
should be checked more carefully.

These issues should be looked at in detail both analytically
and with simulations. However, while simulations
are a useful complement to observations and analytical work in
determining what the structures present in the universe are
actually like, finding out about backreaction with simulations
is not straightforward. Because simulations use
Newtonian gravity and periodic boundary conditions,
backreaction vanishes identically, as discussed in \sec{sec:Buchert}.
Thus the effect of inhomogeneity and anisotropy on the
mean expansion rate would not be visible in simulations even if the
background evolution were dynamically adjusted to take into account
the structures, instead of being predetermined.
For work on the local effects of inhomogeneities and anisotropies
on the expansion in a backreaction context, see \cite{Sicka}.

\paragraph{The backreaction conjecture.}

Let us summarise the physical picture of backreaction-driven
acceleration. As perturbations become non-linear, overdense
regions collapse and form stable structures, and underdense regions
form voids which become ever emptier. The geometry of the
hypersurfaces of constant proper time is no longer everywhere
perturbatively near homogeneous and isotropic, but a fraction of
space is taken up by non-linear structures in various stages of
expansion or collapse. A given proper time no longer corresponds
to a single expansion rate, but to a distribution of expansion rates.
Nevertheless, the description of the expansion in terms of an overall
scale factor remains valid, because the individidual non-linear regions
are small compared to the scales over which we are measuring
observables.
This distribution evolves until it reaches a self-similar regime where
the regions grow at the same rate as the visual horizon. This happens
when the universe is around 10-100 billion years old.

As light rays coming to us pass through hypersurfaces of proper time,
the fraction of space in each stage of expansion (or collapse) that
they encounter changes as new non-linear regions emerge and old
regions evolve.
The conjecture is that the fraction of space in collapsing objects
first grows, diminishing the average expansion rate.
As these regions collapse and their contribution to the
expansion rate is overcome by that of underdense regions,
the average expansion accelerates,
as demonstrated with the toy model in \sec{sec:toy}.
(Note that this mechanism relies on the structures
we know to exist in the universe, as opposed to
speculation of a globally inhomogeneous state \cite{global}.)

This backreaction conjecture should be quantitatively checked
by studying realistic models which can be compared with observations.
However, let us briefly discuss what we can say regarding
compatibility with cosmological observations on the basis of the
Buchert equations and the qualitative description we have outlined.

\subsection{Comparison with observations} \label{sec:obs}

\paragraph{Spatial curvature and the density parameters.}

Acceleration driven by backreaction necessarily involves spatial
curvature. It is instructive to look at the behaviour of the universe
in terms of the density parameters. Dividing \re{Ray} and \re{Ham}
by $3 H^2$, we have \cite{Buchert:1999, Buchert:2003}
\bea
  \label{q} q &\equiv& - \frac{1}{H^2} \frac{\addot}{a} = \frac{1}{2} \Om + 2 \OQ \\
  \label{Omegas} 1 &=& \Om + \OR + \OQ \ ,
\eea

\noindent where $\Om\equiv 8\pi G_N \av{\rho}/(3 H^2)$,
$\OR\equiv-\av{\sR}/(6 H^2)$ and $\OQ\equiv-\sQ/(6 H^2)$ are
the density parameters of matter, spatial curvature and
the backreaction variable, respectively.  As seen from the
definition of $\sQ$ in \re{Q}, the backreaction density parameter
is just minus the relative variance of the expansion rate, plus the
contribution of shear:
$\OQ = - ( \av{\theta^2} - \av{\theta}^2 )/\av{\theta}^2 + 3 \av{\sigma^2}/\av{\theta}^2$.

According to a variety of observations, today we have
$0.15\lesssim\Omn\lesssim0.35$ \cite{Peebles:2004}.
A rough estimate of present-day acceleration from SNIa data is
$-1.2 \lesssim q_0\lesssim-0.3$ \cite{Riess:2004}, though the
lower limit could be extended to at least $-1.6$, and the upper
limit to above zero, so $q_0$ could even be positive today
\cite{Shapiro:2005, Gong:2006}.
Strictly speaking, results from even fairly model-independent analyses
which do not assume the FRW equations but only the FRW metric
\cite{Shapiro:2005, Elgaroy:2006} cannot be directly applied
to backreaction, because of the different behaviour of
the spatial curvature.
In general, the results of SNIa data analysis depend
strongly on the parametrisation used \cite{trans},
and the data seems to contain little model-independent information
about the expansion beyond the fact that it has accelerated in the
recent past \cite{Shapiro:2005, Elgaroy:2006}.
The quoted value of $q_0$ should be understood as a rough
estimate which assumes that the universe has not decelerated
strongly since it started accelerating, sufficient for our purposes.

Given the values of $\Omn, q_0$, the relations \re{q}, \re{Omegas} imply
$0.9\lesssim\ORn\lesssim1.5$, $-0.7\lesssim\OQn\lesssim-0.2$.
It might seem that such a large negative spatial curvature would be
in clear conflict with CMB observations \cite{Spergel:2006}.
However, as with SNIae, CMB analysis is very model- and prior-dependent.
In particular, the analysis leading to tight bounds on spatial
curvature assumes that the equation of state of the negative-pressure
medium which is driving the acceleration does not vary rapidly in time.
In contrast, at the level of the average equations,
backreaction looks like a medium which evolves from the dust
equation of state to having negative pressure.
Such behaviour allows a significant contribution from negative
spatial curvature \cite{Ichikawa:2006}, as the evolution of the
equation of state can cancel the geometrical effect of spatial
curvature on the angular diameter distance.

As with the SNIa data, the non-FRW behaviour of the spatial curvature
means that the CMB analysis in FRW models cannot be applied
to backreaction as is.
Spatial curvature would be expected to be negligible before
backreaction becomes important and then evolve rapidly,
rather than going smoothly like $a^{-2}$, and it could even change sign.
This makes analysis of the CMB anisotropies more involved than
in the FRW case, as the basis functions for slices of constant
curvature cannot be straightforwardly used.
However, if backreaction becomes important only at redshifts of order
unity and below, one might expect the spatially flat FRW framework
to be a good first approximation, as the spatial
curvature becomes significant only on the last leg of
the journey of the CMB from the last scattering surface to us.
At any rate, the issue of light propagation in such an inhomogeneous
and anisotropic spacetime should be studied in the context of a
realistic quantitative model.

Dynamical spatial curvature is a qualitative feature which
is unique to backreaction. It might be useful in distinguishing it
observationally from FRW models, which can share many of the other
features of backreaction, including slowdown before acceleration,
oscillations between acceleration and deceleration and a low Hubble
parameter (see e.g. \cite{Ferrer:2005}).

\paragraph{Large-scale variance.} \label{sec:variance}

It has been argued that backreaction cannot explain the observed
acceleration since the geometry of the universe is so
smooth, i.e. so near the FRW metric \cite{Siegel:2005, Ishibashi:2005}.
However, after non-linear structure formation has started, the geometries
of all regions of the universe are not perturbatively near each other,
as measured by invariant quantities such as the scalar curvature.
For example, the difference in the expansion rate between
expanding and collapsing regions is of order one and their
behaviour is qualitatively different; as we have noted, the
associated breakdown of perturbation theory is well
known in the context of the spherical collapse model.
A perhaps even simpler example is that the space inside
stabilised structures such as galaxies does not expand (or collapse),
so the relative difference between the local expansion rate and
the mean expansion rate is exactly unity.
This difference between the local static metric and the global
expanding metric is also well known, see \cite{Carrera:2006} for discussion
and references, and has recently been studied in the context
of spacetime-dependent couplings \cite{Barrow}.

It is true that despite the presence of non-linear regions,
the density field is smooth when averaged over large scales,
as positive and negative density perturbations cancel due to
conservation of mass. For this reason the energy density of dust
necessarily goes like $\av{\rho}\propto a^{-3}$, as shown by
\re{cons}, and can never lead to acceleration.
However, as we have discussed, it is not the variance of the density
but the variance of the expansion rate which contributes to the
acceleration. As estimated above, a variance of
$(\av{\theta^2}-\av{\theta}^2)/\av{\theta}^2\approx0.2-0.7$
(plus the contribution of shear) today could explain the observations.
The important question is then what is the fraction of space
occupied by regions with non-linear perturbations.
If it is small, the variance can be negligible, whereas if it is
of order one, the expansion will accelerate, unless the effect
of the large variance is overcome by the effect of shear.

Analysis of voids in the two-degree Field Galaxy Redshift Survey found
that 40\% of the survey volume is taken up by voids \cite{Hoyle:2003},
which were required to be very underdense, $\delta\leq-0.9$, and to have
a minimum radius of 10 \mpc. The mean density
contrast of voids was found to be $\delta=-0.94\pm0.02$
and the mean radius was 14.89 $\pm$ 2.65 \mpc.
These numbers are in rough agreement with analysis of earlier
surveys, some of which found larger mean sizes and a bigger
fraction of the volume in voids \cite{voidobs, Hoyle:2001}.
The estimate of the volume fraction is conservative, both because
of the limit on $\delta$ and radius, and because the
void-finding algorithm looks for spherical voids (some of which
are then combined). Simulations have indicated that voids can
have complicated shapes, and that large voids in particular are
typically non-spherical \cite{Shandarin, Shandarin:2005}
(though see also \cite{Sheth:2003, Weygaert:2004}).
The minimum size may also be important.
Analysis of voids in the Millennium simulation (using the \LCDM model)
using the same void-finding algorithm, but with a minimum radius of
6 \mpc, found the mean radius to be 10.45 \mpc, while the mean density
contrast was as low as before, $\delta=-0.92$ \cite{Lee:2006}.
The fact that the mean size is not only smaller than in
\cite{Hoyle:2001, Hoyle:2003}, but near the minimum radius
of those studies suggests that the contribution of small voids may
be important. The prevalence of small voids is also suggested by some
simulations \cite{Colberg:2004}, though not by others \cite{Shandarin}.
In any case, one should be careful when comparing observations of the
real universe with simulations which use the \LCDM model.
It also bears emphasising that there is no generally
agreed definition of a void \cite{Shandarin:2005}, so it is not
straightforward to compare void properties between different studies.

We can make a rough order of magnitude estimate for the lower
limit of the variance implied by the density distribution found
in \cite{Hoyle:2003} by assuming that the volume taken up by voids
is a single homogeneous and isotropic region and the rest of the
universe is another homogeneous and isotropic region.
This is like the toy model in \sec{sec:toy}, but with
region 1 being a general underdense FRW universe instead of being
completely empty. Putting in the numbers $\delta_1=-0.94, v_1=0.4$
and demanding that the mean density equals that of a
spatially flat background gives $\av{\theta^2}/\av{\theta}^2\approx1.08$,
or $\OQn\approx-0.08$. Taking, more consistently, the background
to be the sum of the two regions, we get
$\av{\theta^2}/\av{\theta}^2\approx1.04$, or $\OQn\approx-0.04$.
(In the latter case, we have to take $\Omn=0.45$, as the toy model does not
permit a lower value for the given density distribution.)
These are not far from the the value required by
observations, $\OQn\lesssim-0.2$.
These are likely to be severe underestimates of the variance,
as the essential collapsing regions have been smoothed over.
A realistic estimate should also account for the shear,
which could counter the effect of the variance.

We can also estimate the evolution of the expansion
rate on the basis of observations and \LCDM simulations
of Lyman-$\alpha$ absorbers at redshifts between 2 and 4.5 \cite{Rauch:2005}.
It seems that the Lyman-$\alpha$ absorbers follow the overall Hubble
flow, and allow a fairly direct observation of the expansion rate.
The distribution of expansion rates analysed in \cite{Rauch:2005}
displays the qualitative features which we have identified as crucial
for acceleration.
Most of the volume is underdense and expanding faster than the
mean, and there is a small fraction of space which is collapsing.
The fraction of volume taken up by both the rapidly
expanding and the collapsing regions grows with decreasing redshift, and space
becomes increasingly dominated by voids which expand faster than average.
According to the backreaction conjecture, the volume fraction occupied
by collapsing regions should decrease at low redshifts so that
the mean expansion accelerates. As the redshifts analysed in
\cite{Rauch:2005} do not go below $2$, it is not possible to check
the behaviour in the redshift range where the SNIa data indicate
acceleration. However, we do see from the simulations that the
variance grows with decreasing redshift, implying increasing
backreaction.
At $z=3.8$ we have $\av{\theta^2}/\av{\theta}^2\approx1.05$,
at $z=3.4$ we have $\av{\theta^2}/\av{\theta}^2\approx1.07$,
and at $z=2.0$ we have $\av{\theta^2}/\av{\theta}^2\approx1.26$.
In other words, $\OQ$ evolves from $-0.05$ to $-0.26$, plus the
contribution of shear (which it has not yet been possible
to observationally determine).

Having voids dominate a large fraction of space not only contributes
to the variance of the expansion rate and thus to acceleration, it
is also necessary for boosting $H t$ above the
Einstein--de Sitter value of $2/3$. 
Fitting the parameters of the \LCDM model to the three-year WMAP
data gives $H_0=73.4^{+2.8}_{-3.8}$ km/s/Mpc
and $t_0=13.73^{+0.13}_{-0.17}$ billion years \cite{Spergel:2006},
resulting in (neglecting correlation in the errors)
$H_0 t_0=1.03^{+0.05}_{-0.06}$.
This value is very model-dependent, and we can get a more
robust estimate using the Hubble Space Telescope measurement
of the Hubble parameter, which was recently revised downwards
to $H_0=62.3\pm5.2$ km/s/Mpc \cite{Sandage:2006}
from the old value $H_0=72\pm8$ km/s/Mpc \cite{Freedman:2000}.
Taking $t_0=13\pm1$ billion years, the new value of $H_0$
gives $0.70\lesssim H_0 t_0\lesssim0.97$.

The updated lower value for $H_0$ is easier for
backreaction to accommodate, since $H t\approx1$
requires that almost all of the volume of the universe
is taken up by voids which are almost completely empty.
The volume domination of voids demonstrates what the decreasing
spatial curvature which accompanies backreaction-driven acceleration
implied by the Buchert equations means in physical terms.
Having $H_0 t_0\lesssim1$ is a natural outcome of
the backreaction framework, related to the domination of the
space by voids, whereas in the \LCDM model it is a coincidence.
Given that backreaction cannot raise $H t$ above $1$
(assuming that matter can be treated as dust and vorticity
can be neglected), one can rule out backreaction as an explanation
for the acceleration by showing, in a model-independent manner,
that $H t>1$. (In fact, it is enough to show that the local expansion
rate satisfies the inequality $\theta t/3>1$ somewhere.)

The above estimates show that the qualitative picture of backreaction
outlined on the basis of the Buchert equations and the two-region
toy model is not in obvious disgreement with observations and simulations.
In fact, naive quantitative estimates of the variance of the expansion
rate give the right order of magnitude required for the backreaction
acceleration mechanism to work.
A serious comparison with observations and simulations
will require both a realistic quantitative model for the backreaction 
as well as more careful interpretation of the observations
(since the FRW metric cannot be used) and simulations
(since in backreaction in the Newtonian limit vanishes identically
as an artifact of periodic boundary conditions, and simulations with
a \LCDM background may be misleading).
However, the naive estimates above show that the backreaction
explanation for the acceleration is plausible given what we know
about inhomogeneity and anisotropy in the universe.

\paragraph{Deviations from the average.}

In a homogeneous and isotropic spacetime, it is guaranteed
that a comoving observer will measure the average values for the
expansion rate, energy density and other observables, since there
is no difference between the average and local values.
One would expect this to also hold for spacetimes which are perturbatively
near FRW everywhere. However, in a universe which contains large
inhomogeneities and anisotropies, one can ask what is the relevance
of the averages for an observer making measurements at one particular
location. This is essentially the question of why the approximation of
looking only at an overall scale factor is valid, discussed as
assumption 1 in \sec{sec:FRW}. There are two answers.
The first is that most measurements of cosmological quantities
are made indirectly via observations of the CMB, LSS or SNIa, which
are mostly sensitive to large-scale properties of the universe, rather
than local ones.
The second, more pragmatic, argument is the proven success of the
scale factor ansatz in fitting observations, and the fact that
the size of non-linear structures is small compared to the
visual horizon guarantees the consistency of the ansatz.

However, as the size of structures relative to the visual horizon
is not entirely negligible and the variance of the expansion rate
is large, one could expect to see directional differences in
observables which are sensitive to the large-scale expansion rate
and density distribution, such as the low multipoles of the CMB
and the optical depth.
One can think of this as analysing the statistical scatter around
the average behaviour given by the Buchert equations.
It is tempting to speculate that backreaction could thus link
the observed acceleration to the directional large-angle anomalies
seen in the CMB
\cite{CMBmaps, Hansen:2004, Bielewicz:2005, dipolesyst, Copi:2006}.
In the first-year WMAP temperature map, the southern hemisphere
has an optical depth of $\tau=0.24^{+0.06}_{-0.07}$, while on the northern
hemisphere the optical depth is consistent with zero,
with an upper limit of $\tau<0.08$ (in the frame
which maximises the asymmetry) \cite{Hansen:2004}.
A better studied feature is the presence of a preferred direction
in the low multipoles, which is correlated with the dipole and
the ecliptic plane.
The former suggests an effect related to the structures which
are responsible for our proper motion with respect to the CMB
\cite{CMBmodels2} or systematics regarding the calibration
with the dipole \cite{dipolesyst}, while the latter points towards
a systematical effect associated with the motion of the WMAP satellite.

From cosmology one cannot get a correlation with
the ecliptic, and probably not with the dipole either.
The ecliptic correlation could be a coincidence, as the
low multipoles are still anomalous even if the ecliptic
correlation is neglected, and its significance
has gone down in the three-year WMAP data \cite{Bielewicz:2005, Copi:2006}.
Even neglecting the ecliptic correlation, it is not clear how
backreaction would produce such a distinct preferred direction.
On the other hand, backreaction could tie the directional anomalies
with the observed lack of power at large scales, unlike
most proposals, which explain the directional anomalies by adding a
new contribution which aggravates the amplitude problem.
Since the behaviour of the perturbations is directly related to
the spatial curvature, which is in turn tied to the acceleration,
a connection with directional variation of the Integrated Sachs--Wolfe
effect (and the inferred optical depth) seems plausible.
It is interesting that the template of the anisotropic Bianchi VII$_h$
model fits the anomalies quite well \cite{Bianchi}.
The fitting template corresponds to a homogeneous but anisotropic
universe with shear, vorticity and negative spatial curvature.
Though the details of the Bianchi VII$_h$ universe are different from
what is expected from backreaction, this does show that a globally
inhomogeneous and/or anisotropic universe can explain the anomalies.

Whether any of the CMB directional anomalies are
related to backreaction should be studied in a quantitative model
of light propagation in an inhomogeneous and anisotropic spacetime.
One would also expect to obtain a prediction
for the directional variation of SNIa luminosity distances, which
should be different from that of a linearly perturbed FRW model
\cite{Barausse:2005, Bonvin}, and which might be testable in the
next generation of observations. Such directional variation could be
another clear way, in addition to dynamical spatial curvature, of
distinguishing between acceleration driven by backreaction and
a homogeneous and isotropic medium with negative pressure (or
an equivalent modification of gravity).

\section{Conclusion} \label{sec:con}

\paragraph{Acceleration and inhomogeneity/anisotropy.}

The observational evidence for the acceleration of the universe
is usually interpreted in the framework of
linearly perturbed Friedmann--Robertson--Walker (FRW) models, which
describe a universe that is everywhere almost homogeneous and isotropic.
In the context of such models, a medium with negative
pressure or modified gravity is needed to explain the observations.
This leads to the coincidence problem: why has the exotic matter or
strange gravity become important only recently?
The most significant qualitative change in the universe around the era
where acceleration has been observed is the formation of non-linear
structures, so it seems a natural possibility that the
observed deviation from the general relativistic prediction of the
homogeneous and isotropic cosmological models with normal matter
could be related to the breakdown of homogeneity and isotropy.

The issue of cosmological homogeneity and isotropy has been
extensively discussed over the years by George Ellis and collaborators,
notably in the context of the observational program of cosmology
\cite{Ellis:1975, obscos, hom, Matravers:1995}.
One of the issues they have highlighted is that averaging and applying
the field equations do not commute: in other words, the average properties
of an inhomogeneous and/or anisotropic spacetime do not satisfy the
Einstein equation. The task of finding the model that best describes
the average behaviour of the inhomogeneous universe has been
termed the fitting problem.

The relativistic equations which describe the behaviour of average
quantities in an inhomogeneous and/or anisotropic,
but irrotational, ideal fluid universe have
been derived by Thomas Buchert \cite{Buchert:1999, Buchert:2001}.
The Buchert equations show that it is possible for inhomogeneities
and/or anisotropies to lead to accelerating average expansion in
a dust universe, even though the local acceleration decelerates
everywhere. They also show that the fraction of space occupied by
non-linear regions is the determining quantity, not the size of
the individual regions. Even when the average properties of
space can be described in terms of an overall scale factor,
the evolution of the scale factor does not necessarily follow
the FRW equations.

The possibility that inhomogeneities and/or anisotropies
could lead to acceleration
was studied in the context of linear perturbation theory in
\cite{Rasanen:2003}, and it was suggested that acceleration
could be due to perturbations which have entered the non-linear
regime but haven't yet stabilised. The possibility
of acceleration via backreaction has been numerically verified
\cite{Chuang:2005, Kai:2006, Paranjape:2006}.
However, the physics of how structure formation leads to
acceleration and the question of why acceleration begins much
later than structure formation have been unclear.

We have now clarified these issues, which turn out to be
intimately associated with the process of gravitational collapse.
With a simple toy model, we have explicitly shown how overdense
regions can first slow down the expansion, which then accelerates
as these regions shrink and their contribution to the expansion
rate decreases rapidly as they collapse.

We have also noted that the matter-radiation equality scale imprinted on
the dark matter power spectrum leads to a preferred time
for structure formation that is near the observed
acceleration era. The typical size of collapsing structures
relative to the visual horizon grows rapidly at the start of
structure formation, but then slows down, saturating around
10-100 billion years. A naive look at observations and simulations of
structure in the universe shows that the degree of inhomogeneity
required for backreaction to yield acceleration is plausible.

\paragraph{The backreaction conjecture.}

The backreaction conjecture for the acceleration is simple.
According to the Buchert equations, large
variance of the expansion rate leads to acceleration.
The physical interpretation is simply that the relative volume
taken up by the regions of space which are expanding faster will
come to dominate over the slower expanding regions,
so the average expansion rate will rise.
Collapsing regions, i.e. regions with a negative expansion rate,
give a large contribution to the variance, since they contribute
positively to the mean square but negatively to the square of the mean.
Structure formation involves gravitational collapse, and the
size of the collapsing regions is largest at late times when
acceleration has been observed.

Such an explanation keeps the phenomenological successes of the
FRW scale factor in fitting the observations, while avoiding the
failure of the FRW equations, which has required the introduction
of a medium with negative pressure or modified gravity.
This is in contrast to models which propose explaining the observations
by the effect of inhomogeneities on the propagation of light without
having accelerated expansion \cite{LTBgeo, Biswas:2006}, where
the success of the ansatz that one needs only to look at an overall
scale factor is accidental.

In the context of FRW models, there have been attempts to connect
the late-time acceleration to inflation (via making the same scalar
field responsible for both), the era of matter-radiation equality
(via a tracker field which reacts to the change in the background
equation of state) and dark matter (via unified dark matter and dark energy).
Backreaction involves a subtle link to all these issues. Inflation determines
the initial amplitude of the density perturbations, matter-radiation
equality starts the clock for structure formation, and the nature
of dark matter determines the processed form of the power spectrum
and the time of formation of the first generation of structures.

With many previously unclear conceptual and qualitative issues
settled, the task is now to build a realistic model and make
quantitative estimates that can be compared with observations.
The relevant aspects of observations and simulations
should also be understood better.
On the basis of general considerations we can already state
that we should have $H t<1$, and that there should be observable
amounts of spatial curvature (assuming that vorticity is negligible
and that matter can be treated as dust) \cite{Rasanen:2005}.
There may be a slowdown period preceding the
acceleration, and the expansion may oscillate between
deceleration and acceleration, but these issues have to be worked
out in the context of a detailed model.

Note that there are no new fundamental parameters to adjust, and
any unknowns are due to existing uncertainties about the power
spectrum, the modelling of structure formation and so on:
the backreaction conjecture is eminently falsifiable.
Backreaction analysis simply entails doing the usually
implicit averaging in cosmology in a way that is both
mathematically consistent and takes into account the
structures that are known to be present in the universe,
as has been advocated over the years in the context
of the program of observational cosmology and related work.
Backreaction offers an elegant possible explanation for late-time
acceleration. Whether or not this possibility turns out to be
realised, the effect of structure formation on the expansion rate
should be carefully evaluated to solve the fitting problem and
complete the program of determining the right equations for
describing the overall behaviour of the universe.

\ack

I thank Thomas Buchert for helpful correspondence,
support as well as comments on the manuscript,
several colleagues including John Dubinski
and Constantinos Skordis for clarifying discussions,
and numerous other people for stimulating criticism.
I am also grateful to Michael Rauch for providing 
and explaining the simulation data of \cite{Rauch:2005}.

This paper is dedicated to the victims of operation ``Summer Rain''
and operation ``Just Reward''.\\

\appendix

\setcounter{section}{1}

\end{document}